\documentclass[prc,twocolumn,preprintnumbers,showpacs,superscriptaddress]{revtex4}
\usepackage{amsmath,amssymb,epsfig,bm,color,slashed,
tikz,mathrsfs,graphicx,float,comment,isotope,mathrsfs}
\usepackage{cancel}
\usepackage{aas_macros}
\usepackage[normalem]{ulem}
\usepackage[colorlinks=true,linktocpage=true,linkcolor=blue,citecolor=magenta,urlcolor=blue]{hyperref}

\definecolor{red}{rgb}{0.8,0,0}
\definecolor{violet}{rgb}{0.4,0,0.4}
\definecolor{green}{rgb}{0,0.5,0.0}
\definecolor{navy}{rgb}{0.0,0.0,0.6}
\definecolor{orange}{rgb}{0.8,0.2,0.0}
\usepackage[normalem]{ulem}  % \sout{old text} for strikeout  

\newcommand{\bea}{\begin{eqnarray}}
\newcommand{\eea}{\end{eqnarray}}
\newcommand{\ep}{\epsilon}

\newcommand{\vecp}{{\bm p}}

\newcommand{\Tr}{{\rm Tr}}

\newcommand{\ie}{{\it i.e.}}
\newcommand{\eg}{{\it e.g.}}
\newcommand{\vep}{\varepsilon}
\input{acronym.input}

\begin{document}

\title{Thermoelectric coefficients of two-flavor quark matter from the Kubo formalism}

\author{Harutyun Gabuzyan}
\email{harutyungabuzyan@gmail.com}
\affiliation{Physics Institute, Yerevan State University, Yerevan 0025, Armenia}

\author{Arus Harutyunyan} \email{arus@bao.sci.am}
\affiliation{Byurakan Astrophysical Observatory,
  Byurakan 0213, Armenia}
\affiliation{Physics Institute, Yerevan State University, Yerevan 0025, Armenia}

\author{Armen Sedrakian}
\email{sedrakian@fias.uni-frankfurt.de}
\affiliation{Frankfurt Institute for Advanced Studies, D-60438
  Frankfurt am Main, Germany}
\affiliation{Institute of Theoretical Physics, University of Wroc\l{}aw,
50-204 Wroc\l{}aw, Poland}

\begin{abstract}
  The hot quark matter created in heavy-ion collision experiments can
  exhibit strong temperature and chemical-potential gradients, which in turn can generate electric fields through thermoelectric
  effects. In this work, we investigate two relevant thermoelectric
  coefficients—the thermopower (Seebeck coefficient) and the Thomson
  coefficient—of two-flavor quark matter using the Kubo formalism and
  the Nambu–Jona-Lasinio model as an effective description of dense,
  finite-temperature QCD. The required two-point equilibrium
  correlation functions are evaluated using the Matsubara formalism of
  thermal field theory, applying a $1/N_c$ expansion to the relevant
  multiloop Feynman diagrams. We employ previously derived quark
  spectral functions obtained from one–meson-exchange diagrams above
  the Mott transition temperature. Our numerical results show that
  both thermoelectric coefficients increase approximately linearly
  with temperature and decrease with increasing chemical potential. We
  also estimate the magnitude of the electric fields that can be
  generated in heavy-ion collisions by thermal gradients via the
  Seebeck effect.
\end{abstract}

\maketitle

\section{Introduction}
\label{sec:intro}

Thermoelectric phenomena in hot and dense QCD matter—most notably the
Seebeck, Nernst, and related thermomagnetic transport
coefficients—have attracted increasing interest as probes of
nonequilibrium dynamics in systems with strong temperature gradients
and magnetic fields; for recent reviews
see~\cite{Hattori:2022hyo,Shovkovy:2025yvn}. These coefficients offer
sensitive diagnostics of quark–gluon plasma (QGP) properties, the
evolution of electromagnetic fields, and the interplay between
microscopic scattering processes, momentum-space anisotropies, chiral
structure, and macroscopic hydrodynamic behavior. Their relevance
extends from relativistic heavy-ion collisions, where intense
short-lived electromagnetic fields and steep spatial gradients can
induce substantial charge and heat transport, to astrophysical
environments such as neutron stars, core-collapse supernovae, and
binary-merger remnants, where dense quark matter, strong fields, and large thermodynamic gradients naturally occur. Over the past decade, a broad range of theoretical frameworks—including kinetic theory,
effective field theory, and holography—have been employed to characterize these transport properties across both hadronic and partonic phases of strongly interacting matter.

Strong-coupling studies based on holographic models provided some of
the earliest insights into thermoelectric conductivities, with
analyses such as~\cite{Kim2015} offering benchmarks for Seebeck and
Nernst coefficients at finite magnetic field. Complementary
kinetic-theory calculations in hadronic matter followed, including studies of the Seebeck effect in hadron resonance gas
models~\cite{Bhatt2019,Das2020}. Thermoelectric response in QCD matter
has since been explored in a variety of magnetic-field regimes. Works
such as~\cite{Dey2020,Dey2021} analyzed Seebeck and related
coefficients in quantizing and weak magnetic fields, while the impact
of collisional dynamics was studied in~\cite{Kurian2021} using
relaxation-time and BGK-type approaches. The role of plasma anisotropy
was emphasized in~\cite{Zhang2021,Rath2025}, which demonstrated that both
Seebeck and Nernst coefficients are highly sensitive to momentum-space
deformations in the QGP. More recent developments updated perturbative
QCD treatments of the Seebeck effect~\cite{Shaikh2025}, and analyses
incorporating medium-dependent quark masses in anisotropic
plasmas~\cite{Khan2024}. Time-dependent magnetic fields--an essential feature of heavy-ion collisions--have been addressed in studies of both
hadronic and partonic matter~\cite{Kumar2025,Singh2024,Singh2025},
highlighting the dynamical interplay between evolving fields and
induced thermoelectric currents.

Effective model approaches have provided additional perspectives on transport in quark matter. For example, Nambu–Jona-Lasinio– (NJL) based  studies such as~\cite{Abhishek2020,Zhang2022} examined thermoelectric coefficients using relaxation times derived from meson-exchange scattering, yielding a complementary quasiparticle description within strongly interacting matter. Two principal theoretical approaches are available for obtaining
transport coefficients from an underlying microscopic description. For
weakly interacting systems, kinetic theory based on the Boltzmann equation provides a natural framework for computing transport in terms
of quasiparticle scattering processes. In contrast, strongly
interacting quantum systems require a fully quantum-statistical
treatment, where transport coefficients are related to appropriate
equilibrium correlation functions via the Kubo formulas. The latter
approach, grounded in the Liouville equation for the nonequilibrium density matrix, is applicable in principle to systems with strong
interactions and collective dynamics. These two approaches are complementary 
under certain conditions, as for example, in a weakly coupled
relativistic scalar field theory, where sums of an infinite
class of diagrams lead to linear integral equation, which
are equivalent to a linearized Boltzmann equation~\cite{Jeon1995}.

In this work, we employ the Kubo formalism to compute the thermal
conductivity and thermoelectric power of hot quark
matter. Specifically, thermoelectric power plays a crucial role when
temperature gradients and electric fields coexist, as in the early
stages of heavy-ion collisions. We adopt the NJL
model, an effective chiral theory where quark–quark interactions are
encoded through four-fermion contact terms. The nonequilibrium density matrix is constructed by perturbing the grand canonical
ensemble and solving the Liouville equation to linear order in the
applied electric field and temperature gradient. The relevant two-point correlation functions of electric and heat currents are
evaluated in the Matsubara (imaginary-time) formalism, which, at
leading order, are given by single-loop diagrams with fully
dressed quark propagators, while vertex corrections are suppressed in $1/N_c$ counting scheme.
This type of approach has been applied in a broader context beyond the NJL model. For example, shear viscosity in field theories is dominated by one-loop diagrams with finite thermal widths, while higher-order loops are negligible~\cite{Jeon1993,Ghosh2014}. Similarly, electrical conductivities of baryon–meson systems have been computed using current–current correlators with thermal widths determined from in-medium scattering~\cite{Ghosh2017}. We emphasize that, in the limit of vanishing widths, the correlators diverge and the transport coefficients vanish.

This work is organized as follows. Section~\ref{sec:Kubo} derives the
Kubo formulas for the thermoelectric transport coefficients starting
from the quantum Liouville equation for the density
matrix. Section~\ref{sec:NJL} computes the relevant two-point
correlation functions and expresses the transport coefficients in
terms of the quark spectral function. In Sec.~\ref{sec:results} we
provide numerical results for the thermal conductivity and
thermopower. Section~\ref{sec:summary} gives a brief summary of our
results.  Appendix~\ref{app:Green_func} provides some details on the
derivation of the Kubo formulas.  We use the natural units with
$\hbar= c = k_B = 1$, $e=\sqrt{4\pi\alpha}$, $\alpha=1/137$, and the
metric signature $(1,-1,-1,-1)$.

\section{Derivation of the Kubo formulas}
\label{sec:Kubo}

The Kubo formalism relates the transport coefficients of a statistical ensemble to certain equilibrium
correlation functions, which in turn can be computed using equilibrium many-body
techniques~\cite{1957JPSJ...12..570K,1957JPSJ...12.1203K,Mahan}. The thermodynamic state of a macroscopic quantum system can be described by means of the time-dependent density matrix $\rho(t)$. The density matrix 
(grand canonical distribution)
%----------------------------------------
\bea\label{eq:rho_0}
\rho_0=e^{\Omega_0 -\beta K_0},\quad K_0=\mathcal{H}_0-\mu \mathcal{N}
\eea
%----------------------------------------
applies to the equilibrium system with unperturbed Hamiltonian
$\mathcal{H}_0$ and conserved particle number $\mathcal{N}$, where
$\beta$ is the inverse temperature, $\mu$ is the chemical potential,
and $\Omega_0=-\ln\left(\Tr e^{-\beta K_0}\right)$ is the grand
thermodynamic potential.  We will assume that the system is described
by the equilibrium density matrix at the initial point of time, which we take at $t=-\infty$.

The derivation of the Kubo formulas for thermoelectric transport
coefficients proceeds as follows. In a nonequilibrium state, the
density matrix becomes time dependent and obeys the Liouville
equation, which in the Schr\"odinger representation reads
%----------------------------------------
\bea\label{eq:Liouville}
\frac{\partial\rho(t)}{\partial t}=
-i[\mathcal{H}, \rho(t)],
\eea
%----------------------------------------
where $\mathcal{H}$ is the Hamiltonian of the system including the
external perturbation. In this representation, the operators acting on the quantum states of the system are time independent; thus, the
explicit time dependence of $\rho(t)$ arises solely from the time
dependence of external fields and thermodynamic parameters. In what
follows, we use the subscript $H$ to denote operators in the
Heisenberg representation, while operators without a subscript are
understood to be in the Schr\"odinger representation.

Let us now consider a situation in which thermal and chemical
gradients and an electrical field are present simultaneously.  If these
are sufficiently small, then a so-called local-equilibrium density
matrix can be constructed. To do this, it is convenient to write the
grand canonical distribution in terms of the energy-momentum tensor
and the particle current density using the relations
%---------------------------------
\bea\label{eq:H_N}
{\cal H}_0 =\int d\bm r\, {T}^{00}(\bm r),\qquad
{\cal N}=\int d\bm r\, {N}^{0}(\bm r).
\eea
%---------------------------------
Substituting these into Eq.~\eqref{eq:rho_0} we obtain
%---------------------------------
\bea\label{eq:Gibbs_inv}
{\rho}_{0} = \exp\bigg\{\Omega_0 -\int d\bm r  \Big[\beta {T}^{00}(\bm r)-\alpha {N}^{0}(\bm r)\Big]\bigg\},
\eea
%--------------------------------
with $\alpha=\beta\mu$.
The local-equilibrium  density matrix can now be constructed 
by replacing $\beta\to\beta(\bm r,t)$, $\alpha\to\alpha(\bm r,t)$, which gives
%----------------------------------------
\bea\label{eq:rho_l}
\rho_{l}(t)&=&e^{a(t)},\\
a(t)&=&\Omega_{l}(t)-\int d\bm r\Big[\beta(\bm r,t) T^{00}(\bm
r)-\alpha(\bm r,t)N^0(\bm r)\Big],
\nonumber\\
\eea
%----------------------------------------
where the local-equilibrium grand thermodynamic potential is given by 
%----------------------------------------
\bea\label{eq:Omega_l}
\Omega_l(t) &=&-\ln\Tr \exp\Bigg\{-\int d\bm r \nonumber\\
&\times&\Big[\beta(\bm r,t)T^{00}(\bm r)
-\alpha(\bm r,t)N^0(\bm r)\Big]\Bigg\}.
\eea
%----------------------------------------
One should keep in mind, however, that the local equilibrium density matrix
$\eqref{eq:rho_l}$ does not satisfy the
Liouville equation $\eqref{eq:Liouville}$ and therefore cannot
describe irreversible processes. However, it can serve as a starting point (\ie, the zeroth-order approximation) to construct an exact density matrix at the required order in thermodynamic perturbations, such as external fields and thermal gradients. In this work, we are interested only in linear response of the system to thermoelectric perturbations. The derivation of the nonequilibrium density matrix up to the second order in perturbations in a full relativistic framework was carried out recently in Refs.~\cite{Harutyunyan2022,Harutyunyan2025}.

Assuming weak electric fields and small deviations from thermal
equilibrium, we seek the exact density matrix in the form
$\rho(t) = \rho_{l}(t) + f(t)$, where $f(t)$ is a small
perturbation. Then we obtain
%----------------------------------------
\bea\label{eq:df_dt1}
i\frac{\partial f}{\partial t}=[{\cal H},f] + [{\cal H},\rho_{l}]
-i\frac{\partial\rho_l}{\partial t}.
\eea
%----------------------------------------
Note that the local equilibrium density matrix $\rho_l$ does not
commute with the Hamiltonian, because of the nonuniformity of
temperature and chemical potential, as well as the presence of external electric field (in equilibrium we would have
$[{\cal H}_0,\rho_{0}]=0$). Thus, it can be anticipated that the
commutator $[{\cal H},\rho_{l}]$ should be proportional to the
gradients of these quantities and the electric field, as shown below.
The detailed procedure of solving Eq.~\eqref{eq:df_dt1} is given in
Appendix~\ref{app:Liouville}. The result reads
%----------------------------------------
\bea\label{eq:ft_final_main}
f(t)=ie^{-i{\cal H}t}\left\{\int_{-\infty}^t\,dt' \Big[V(t')+W(t'),\rho_0\Big]\right\}e^{i{\cal H}t},
\eea
%----------------------------------------
where $V(t')$ and $W(t')$ stand for electric and thermal
perturbations, respectively, and are given by
%----------------------------------------
\bea\label{eq:V_pert}
V(t) &=& -\int_{-\infty}^{t} dt'' \! \int d\bm r'\,J^k(\bm r',t'')
E_{k}(\bm r',t''),\\ 
\label{eq:W_pert}
W(t) &=& \frac{1}{\beta} \int_{-\infty}^{t} dt'' \! \int d\bm r'\,H^k(\bm r',t'')\partial_k\beta(\bm r',t'').\quad
\eea
%----------------------------------------
Now we are in a position to evaluate the statistical averages of the
electrical and heat currents according to
%----------------------------------------
\bea\label{eq:j_h_av}
j^i(\bm r,t) =\Tr\{f(t) J^i(\bm r)\},~~
h^i(\bm r,t) =\Tr\{f(t) H^i(\bm r)\},\,
\eea
%----------------------------------------
where we took into account that there are no currents in local
equilibrium, \ie, $\Tr(\rho_{l}J^i)=0$, $\Tr(\rho_{l}H^i)=0$.
Substituting Eq.~\eqref{eq:ft_final_main} into Eq.~\eqref{eq:j_h_av}
and using the cyclic properties of the trace we obtain
%----------------------------------------
\bea
j^i(\bm r,t)
&=&i\Tr\left\{\bigg(\int_{-\infty}^t\,dt' \big[V(t')+W(t'),
  \rho_0\big]\bigg)J^i(\bm r,t)\right\}.
\nonumber\\
\eea
%----------------------------------------
Here $J^i(\bm r,t)$ is the current operator in the Heisenberg representation. Again, using the cyclic properties of the trace, the operators under the trace can be rearranged as follows
%----------------------------------------
\bea
\Tr\Big\{\big[V(t'),\rho_0\big]J^i(\bm r,t)\Big\} &=& \Tr\Big\{\rho_0\big[J^
i(\bm r,t),V(t')\big]\Big\}\nonumber\\
&=&\langle\big[J^i(\bm r,t),V(t')\big]\rangle,
\eea
%----------------------------------------
where the angle brackets mean a trace over the thermal equilibrium distribution at nonzero temperatures.
Thus, the final results for the expectation values of the electric and heat currents read
%----------------------------------------
\bea\label{eq:j_av}
j^i(\bm r, t) &=&
i\!\int_{-\infty}^tdt'\langle \big[J^i(\bm
r,t),V(t')\big]\rangle\nonumber\\
&+& i\!\int_{-\infty}^tdt'\langle \big[J^i(\bm r,t),W(t')\big]\rangle,\\
\label{eq:h_av}
h^i(\bm r, t) &=&
i\!\int_{-\infty}^tdt'\langle \big[H^i(\bm r,t),V(t')\big] \rangle
\nonumber\\
&+& i\!\int_{-\infty}^tdt'\langle \big[H^i(\bm r,t),W(t')\big] \rangle.
\eea
%----------------------------------------
Next, we substitute the expressions for $V(t)$ and $W(t)$ given by Eqs.~\eqref{eq:V_pert} and \eqref{eq:W_pert}, assuming homogeneous electric and thermal fields with monochromatic time dependence. Furthermore, we will assume that the perturbations are switched on adiabatically from the moment $t\to -\infty$, thus taking the perturbations in the form $E_{k}(\bm r,t)= E_k e^{(s-i\omega) t}$, $\partial_{k}\beta(\bm r,t)= \partial_{k}\beta e^{(s-i\omega) t}$ with $s\to +0$, where $E_k, \partial_k\beta={\rm const}$. Then we obtain
% ----------------------------------------
\begin{widetext}
\bea\label{eq:j_i}
j^i(\bm r,t)
&=& -i\int_{-\infty}^t dt'\!\int_{-\infty}^{t'} dt''e^{i\omega^\star (t-t'')}\!\int d\bm r'\langle\big[J^i(\bm r,t),J^k(\bm r',t'')\big]\rangle E_k(\bm r,t)\nonumber\\
&+& \frac{i}{\beta}\int_{-\infty}^t dt'\!\int_{-\infty}^{t'} dt''e^{i\omega^\star (t-t'')}\!\int d\bm r'\langle\big[J^i(\bm r,t),H^k(\bm r', t'')\big]\rangle\partial_k\beta(\bm r,t),\\
%-------------------------
\label{eq:h_i}
h^i(\bm r,t)
&=& -i\int_{-\infty}^t dt'\!\int_{-\infty}^{t'} dt''e^{i\omega^\star (t-t'')}\!\int d\bm r'\langle\big[H^i(\bm r,t),J^k(\bm r',t'')\big]\rangle E_k(\bm r,t)\nonumber\\
&+&\frac{i}{\beta}\int_{-\infty}^t dt'\!\int_{-\infty}^{t'} dt''e^{i\omega^\star (t-t'')}\!\int d\bm r'\langle\big[H^i(\bm r,t),H^k(\bm r', t'')\big]\rangle\partial_k\beta(\bm r,t),\qquad
\eea
%----------------------------------------
with $\omega^\star=\omega+i0$.  The inner integrands in these formulas
contain the two-point retarded Green's functions in the coordinate space in a thermal medium.

Defining now the transport coefficients by
%----------------------------------------
\bea\label{eq:sigma_ik}
\sigma^{ik}(\omega) &=& i\int_{-\infty}^t dt'\!\int_{-\infty}^{t'} dt''e^{i\omega^\star (t-t'')}\!\int d\bm r'\langle\big[J^i(\bm r,t),J^k(\bm r',t'')\big]\rangle,\\
\label{eq:gamma_ik}
\gamma^{ik}(\omega) &=& i\int_{-\infty}^t dt'\!\int_{-\infty}^{t'} dt''e^{i\omega^\star (t-t'')}\!\int d\bm r'\langle\big[H^i(\bm r,t),J^k(\bm r',t'')\big]\rangle,\\
\label{eq:alpha_ik}
\alpha^{ik}(\omega) &=& i{\beta}\int_{-\infty}^t dt'\!\int_{-\infty}^{t'} dt''e^{i\omega^\star (t-t'')}\!\int d\bm r'\langle\big[J^i(\bm r,t),H^k(\bm r', t'')\big]\rangle,\\
\label{eq:kappat_ik}
\tilde\kappa^{ik}(\omega) &=& i{\beta}\int_{-\infty}^t dt'\!\int_{-\infty}^{t'} dt''e^{i\omega^\star (t-t'')}\!\int d\bm r'\langle\big[H^i(\bm r,t),H^k(\bm r', t'')\big]\rangle,
\eea
%----------------------------------------
we can write Eqs.~\eqref{eq:j_i} and \eqref{eq:h_i} in the form
%----------------------------------------
\bea\label{eq:jh_1}
j^i= -\sigma^{ik} E_{k}
+\beta^{-2}\alpha^{ik} \partial_k\beta,\qquad
h^i= -\gamma^{ik} E_{k} +\beta^{-2}\tilde{\kappa}^{ik}
\partial_k\beta.
\eea
% ----------------------------------------
\end{widetext}
In an isotropic medium, these tensors become diagonal, \ie,
$\sigma^{ik}=\delta^{ik}\sigma$, etc., therefore, substituting also
$\beta=T^{-1}$, we can write Eq.~\eqref{eq:jh_1} in the form (recall
that the components of the vectors $\bm j$, $\bm h$ and $\bm E$ are
$j^i=-j_i$, etc.)
%----------------------------------------
\bea\label{eq:jh_2}
\bm j= \sigma \bm E -\alpha \bm{\nabla}  T,\qquad
\bm h= \gamma \bm E -\tilde{\kappa}\bm{\nabla}  T.
\eea
%----------------------------------------

Now we express $\bm h$ in terms of $\bm{\nabla}  T$ and $\bm j$ and define the thermoelectric power (thermopower) $Q=\alpha/\sigma =\gamma/T\sigma$, and thermal conductivity 
$\kappa =\tilde{\kappa}-TQ^2\sigma$ (the relation $\gamma=\alpha T$ follows from the symmetry property of the Green's function~\cite{1957JPSJ...12..570K,Harutyunyan2022,Harutyunyan2025}). Finally, in the simultaneous presence of electric field $\bm E$ and thermal gradient $\bm{\nabla}  T$, the statistical averages of the electrical and thermal currents
in linear response theory can be written in the form
%----------------------------------------
\bea\label{eq:currents_reversed}
\bm j = \sigma\bm E 
-\sigma Q\bm{\nabla}  T,\qquad
\bm h = -\kappa\bm{\nabla}  T + TQ\bm j.
\eea
%----------------------------------------

Performing the temporal integrations in the expressions~\eqref{eq:sigma_ik}--\eqref{eq:kappat_ik}, the Kubo formulas for three thermoelectric coefficients can be derived, see Appendix~\ref{app:Green_func} for details,
%----------------------------------------
\bea\label{eq:sigma}
\sigma &=& -\frac{d}{d\omega} {\rm Im}\Pi^{R}_\sigma (\omega)\bigg|_{\omega =0},\\
\label{eq:kappa_til}
\tilde{\kappa} &=& -\beta\frac{d}{d\omega} {\rm Im}\Pi^{R}_{\tilde{\kappa}} (\omega)\bigg|_{\omega =0},\\
\label{eq:Qsigma}
\sigma Q &=& -\beta\frac{d}{d\omega} {\rm Im}\Pi^{R}_{Q} (\omega)\bigg|_{\omega =0},
\eea
%----------------------------------------
where the two-point retarded Green's functions on the right-hand sides are defined as
%----------------------------------------
\bea\label{eq:corkappasigma1}
\Pi^{R}_\sigma(\omega) &=& -i\int_{0}^{\infty}dt\!
e^{i\omega t}\!\!\int d\bm r\langle
\big[J_{1}(\bm r,t),J_{1}(0)\big]\rangle_{0},\\
\label{eq:corkappasigma2}
\Pi^{R}_{\tilde{\kappa}}(\omega) &=& -i\int_{0}^{\infty}dt\!
 e^{i\omega t}\!\!\int d\bm r\langle
\big[H_{1}(\bm r,t),H_{1}(0)\big]\rangle_{0},\\
\label{eq:corkappasigma3}
\Pi^{R}_Q(\omega) &=& -i\int_{0}^{\infty}dt\!
e^{i\omega t}\!\!\int d\bm r\langle
\big[J_{1}(\bm r,t),H_{1}(0)\big]\rangle_{0}.\quad
\eea
%----------------------------------------
Equations~\eqref{eq:sigma}--\eqref{eq:corkappasigma3} are known as Kubo
formulas~\cite{1957JPSJ...12..570K,1957JPSJ...12.1203K,Mahan}. These equations apply to an arbitrary quantum-statistical ensemble without restrictions on the strength of the couplings of the underlying theory. In the following, we will derive more specific expressions suitable for the NJL model with contact scalar and pseudoscalar couplings among quarks by applying the $1/N_c$ expansion to select the dominant diagrams contributing to the correlation functions.

The relations~\eqref{eq:currents_reversed} lead to several
thermoelectric effects, among which we will discuss the Seebeck effect
and the Thomson effect (see, \eg, Refs~\cite{Ziman1979,Landau8}).  The
Seebeck effect refers to the generation of an electric field in a
conducting material due to a temperature gradient in the absence of an
electric current: $\bm{E} = Q \bm{\nabla} T$. The ``Seebeck coefficient" $Q$
is defined as the proportionality between $\bm{\nabla} T$ and $\bm{E}$ and
is identical to the thermopower.

The Thomson effect is observed in the case where both electric current
and heat flux are present simultaneously. To analyze the effect, let
us calculate the total heat ${\cal \bm Q}$ released per unit time per unit volume of the system. The heat released due to thermal conduction is equal to $-{\rm div} \bm h$, and the heat released due to electric conduction is $\bm j\cdot\bm E$, therefore
%----------------------------------------
\bea\label{eq:heat_released}
{\cal \bm Q}=-{\rm div} \bm h +\bm j\cdot\bm E={\rm div}(\kappa\bm{\nabla}  T)
+\frac{\bm j^2}{\sigma} 
-T\bm j\cdot\bm{\nabla} Q,
\eea
%----------------------------------------
where we used Eq.~\eqref{eq:currents_reversed} and the charge conservation ${\rm div}\bm j=0$.
The first term here is associated with pure thermal conductivity, and the second term is referred to as joule heat. The third term is of particular interest, as it encompasses the specific thermoelectric effects.

Since our focus here is hot quark matter created in heavy-ion collisions, it makes sense to consider thermoelectric effects at constant baryon density, which is one-third of the quark number density $n$. 
Then the change in the thermopower $Q$ is related only to the temperature gradient, therefore the heat released due to the simultaneous presence of electric and heat fluxes (the Thomson effect) will be $\rho\bm j\cdot\bm{\nabla}  T$, where 
%----------------------------------------
\bea\label{eq:Thomson}
\rho=-T\left(\frac{dQ}{dT}\right)_n
\eea
%----------------------------------------
is called the ``Thomson coefficient".  Note that the Thomson heat
changes sign when the direction of the current is reversed, so the Thomson coefficient can be either positive or negative. For $\rho>0$,
heat is released when the current flows toward higher temperatures and
absorbed when it flows in the opposite direction. For $\rho<0$, the
situation is reversed.

\section{Correlation functions in the NJL model}
\label{sec:NJL}

We consider two-flavor quark matter described by the NJL Lagrangian~\cite{1961PhRv..124..246N,1961PhRv..122..345N,1991PrPNP..27..195V,1992RvMP...64..649K,2005PhR...407..205B}
%----------------------------------------
\bea\label{eq:lagrangian} \mathcal{L}=\bar\psi(i\slashed
\partial-m_0)\psi+ \frac{G}{2}\left[(\bar\psi\psi)^2+ (\bar\psi
  i\gamma_5\bm\tau\psi)^2\right],
\eea%----------------------------------------
where $\psi=(u,d)^T$ is the isodoublet quark field, $m_0=5.5$ MeV is
the current quark mass, $G=10.1$ GeV$^{-2}$ is the effective
four-fermion coupling constant, and $\bm\tau$ is the vector of Pauli
isospin matrices. This Lagrangian describes four-fermion
scalar-isoscalar and pseudoscalar-isovector interactions between
quarks with the corresponding bare vertices $\Gamma^0_{s}=1$ and
$\Gamma^0_{ps}=i\bm\tau\gamma_5$.  Four-fermion contact interactions
make the model nonrenormalizable; therefore, a 3-momentum cutoff
($p\le\Lambda$) should be applied to regularize the ultraviolet
divergences. The cutoff parameter $\Lambda$, the current quark masses
$m_0$, and the four-fermion coupling $G$ constitute the free
parameters of the model. Their values are chosen to reproduce
experimental values of meson masses and the pion decay constant.
Below, we will adopt the value $\Lambda=0.65$ GeV. 

The symmetrized energy-momentum tensor of the model is given by
%----------------------------------------
\bea\label{eq:energymom}
T_{\mu\nu}=\frac{i}{2}(\bar\psi\gamma_{\mu}
\partial_{\nu}\psi +\bar\psi\gamma_{\nu}
\partial_{\mu}\psi)-g_{\mu\nu}\mathcal{L},
\eea
%----------------------------------------
and the quark number and charge currents are defined as
%----------------------------------------
\bea\label{eq:current}
N_{\mu}=\bar\psi\gamma_{\mu}\psi,\qquad
J_{\mu}=\bar\psi\hat{q}\gamma_{\mu}\psi,
\eea
%----------------------------------------
where 
%----------------------------------------
\bea\label{eq:charge}
\hat{q}=e 
\begin{pmatrix}
    2/3 & 0 \\
    0 & -1/3 
\end{pmatrix}
\eea%---------------------------------------- 
is the charge matrix in flavor space, with $e$ being the elementary
charge.

Next, we evaluate the correlation functions given by
Eqs.~\eqref{eq:corkappasigma1}--\eqref{eq:corkappasigma3} in thermal
equilibrium using the imaginary-time Matsubara technique (see, \eg,
Ref.~\cite{Mahan} for details). We mainly follow the derivations of
Ref.~\cite{Harutyunyan2017}, but consider also the thermopower and thermoelectric correction to the thermal conductivity, which were omitted there.

The corresponding Matsubara correlation functions read
%----------------------------------------
\bea\label{eq:corsigma_Q_m}
\Pi^{M}_{\sigma}(\omega_n) &=& \Pi^M_{JJ}(\omega_n),\quad 
\Pi^{M}_{Q}(\omega_n)=\Pi^M_{JT}(\omega_n)-h\Pi^M_{JN}(\omega_n),\nonumber\\\\
\label{eq:corkappa_m}
\Pi^{M}_{\tilde{\kappa}}(\omega_n) &=& \Pi^M_{TT}(\omega_n)-2h\Pi^M_{TN}(\omega_n)+h^2\Pi^M_{NN}(\omega_n),
\eea
%----------------------------------------
with
%----------------------------------------
\bea\label{eq:corTT_m}
-\Pi^{M}_{TT}(\omega_n) &=& \int_{0}^{\beta}d\tau 
e^{i\omega_n\tau}\!\!\int d\bm r\langle
{\cal T}_{\tau}\big\{T_{10}(\bm r,\tau),T_{10}(0)\big\}\rangle_0, \nonumber\\\\
\label{eq:corTN_m}
-\Pi^{M}_{TN}(\omega_n)&=& \int_{0}^{\beta}d\tau 
e^{i\omega_n\tau}\!\!\int d\bm r\langle
{\cal T}_{\tau}\big\{T_{10}(\bm r,\tau),N_{1}(0)\big\}\rangle_0, \nonumber\\\\
\label{eq:corJT_m}
-\Pi^{M}_{JT}(\omega_n)&=& \int_{0}^{\beta}d\tau 
e^{i\omega_n\tau}\!\!\int d\bm r\langle
{\cal T}_{\tau}\big\{J_{1}(\bm r,\tau),T_{10}(0)\big\}\rangle_0, \nonumber\\\\
\label{eq:corNN_m}
-\Pi^{M}_{NN}(\omega_n)&=&\int_{0}^{\beta}d\tau 
e^{i\omega_n\tau}\!\!\int d\bm r\langle
{\cal T}_{\tau}\big\{N_{1}(\bm r,\tau),N_{1}(0)\big\}\rangle_0, \nonumber\\\\
\label{eq:corJJ_m}
-\Pi^{M}_{JJ}(\omega_n)&=&
\int_{0}^{\beta} d\tau
 e^{i\omega_n\tau} \int d\bm r\langle
{\cal T}_{\tau}\big\{J_1(\bm r,\tau),J_1(0)\big\}\rangle_0, \nonumber\\\\
\label{eq:corJN_m}
-\Pi^{M}_{JN}(\omega_n)&=& \int_{0}^{\beta}d\tau 
e^{i\omega_n\tau}\!\!\int d\bm r\langle
{\cal T}_{\tau}\big\{J_{1}(\bm r,\tau),N_{1}(0)\big\}\rangle_0, \nonumber\\
\eea
%----------------------------------------
where $T_{10}(\bm r,\tau)$, $N_1(\bm r,\tau)$, and $J_1(\bm r,\tau)$
are obtained from $T_{10}(\bm r,t)$, $N_1(\bm r,t)$, and $J_1(\bm r,t)$
via the Wick rotation $t\to -i\tau$, and ${\cal T}_\tau$ is the time-ordering
operator for imaginary time $\tau$.
In Eqs.~\eqref{eq:corsigma_Q_m} and \eqref{eq:corkappa_m} we substituted the operator of heat flux~\eqref{eq:heat_current} and used the symmetry of the
correlation function with respect to its arguments~\cite{1957JPSJ...12..570K,Harutyunyan2017}. The required retarded
correlation functions \eqref{eq:corkappasigma1}--\eqref{eq:corkappasigma3} can be obtained from their Matsubara counterparts \eqref{eq:corsigma_Q_m} and \eqref{eq:corkappa_m} by an analytic continuation
$i\omega_n\to \omega +i\delta$.  Note that the transformation to
imaginary time implies a change of the derivative
$\partial_0\to i\partial_\tau$.  Because $T_{\mu\nu}$, $N_\mu$, and $J_\mu$ are bosonic operators, the Matsubara
frequencies assume even integer values $\omega_n=2\pi nT$,
$n=0,\pm 1,\ldots$. The $T_{10}$ component of the energy-momentum
tensor~\eqref{eq:energymom} in the Matsubara frequency space is given by
%----------------------------------------
\bea\label{eq:to1m}
T_{10}(\bm r,\tau)&=&i\bar\psi(\bm
r,\tau)\frac{\gamma_0}{2}\partial_1\psi(\bm r,\tau)
\nonumber\\
&+&i\bar\psi(\bm r,\tau)\frac{\gamma_1}{2}i\partial_\tau\psi(\bm r,\tau).
\eea
%

%--------------------------------------------------------------
\begin{figure}[hbt] 
\begin{center}
\includegraphics[width=6cm,keepaspectratio]{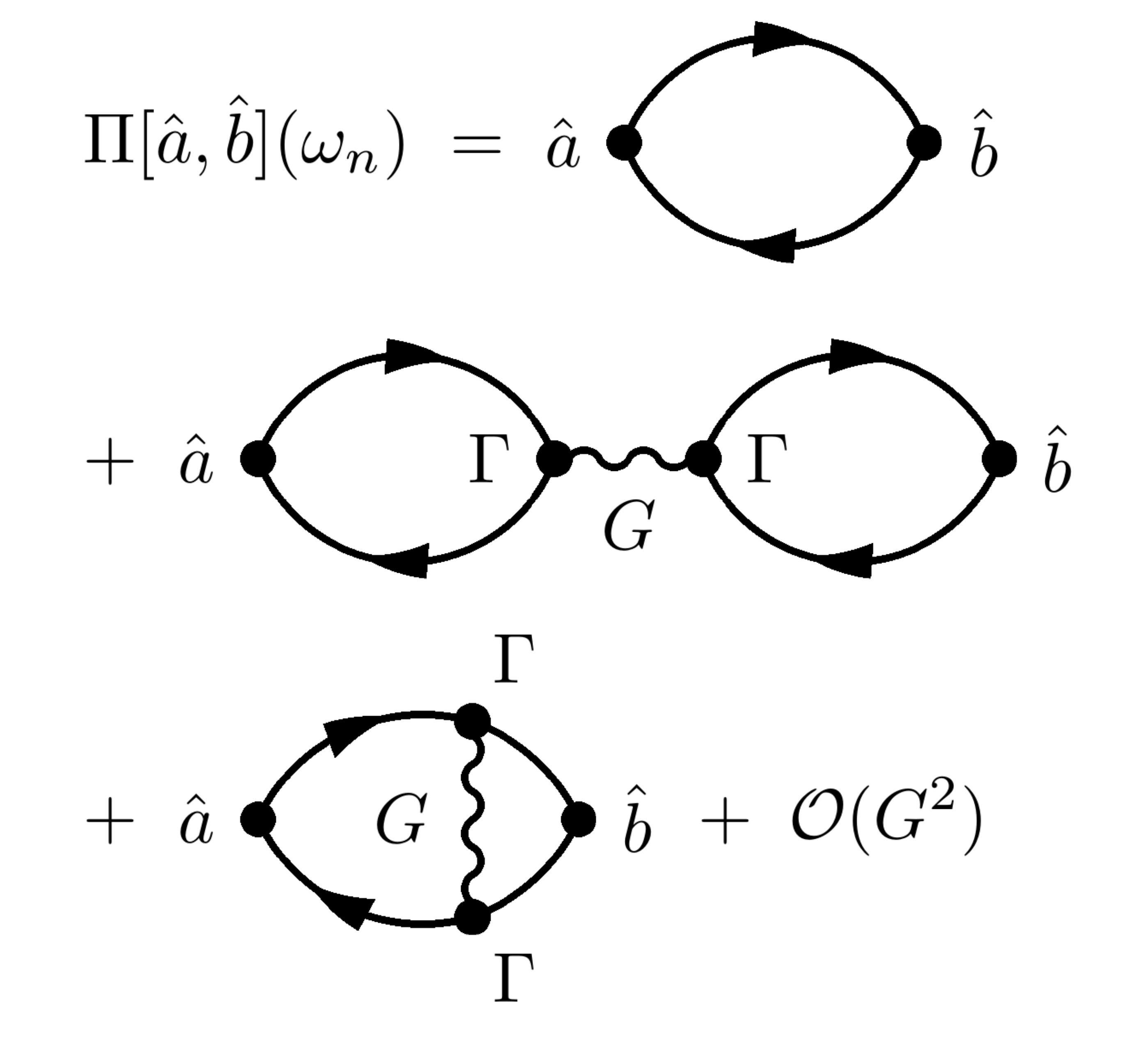}
\caption{ Contributions to a generic two-point correlation function
  from ${\cal O}(N_c^1)$ (first and second lines) and
  ${\cal O}(N_c^0)$ (the third line) diagrams up to the first-order
  terms with respect to the coupling constant $G$ associated with a
  pair of $\Gamma=\Gamma^0_{s/ps}$ matrices. The straight lines
  represent dressed quark propagators. }
\label{fig:loops} 
\end{center}
\end{figure}
%--------------------------------------------------------------
Each of the correlation functions above can be represented as an
infinite sum of diagrams shown schematically in Fig.~\ref{fig:loops}
[here the operators $\hat{a}$ and $\hat{b}$ contain momentum and
  $\gamma$ matrices; their form for each correlation function should
  be specified be means of Eqs.~\eqref{eq:current} and
  \eqref{eq:to1m}].  The propagators here are dressed, \ie, the
dispersive effects are taken into account in their spectral width.  In
the $1/N_c$ expansion scheme, where $N_c$ is the number of colors, the
diagrams in the first and second lines are of order ${\cal O}(N_c^1)$,
whereas the diagram in the third line is of order ${\cal
  O}(N_c^0)$. Following the argument of Ref.~\cite{Harutyunyan2017}
(see also
Refs.~\cite{1994PhRvC..49.3283Q,2008JPhG...35c5003I,2008EPJA...38...97A,LW14,LKW15,2016PhRvC..93d5205G,LangDiss},
we retain only the first-line diagram, as the others shown, along with
higher-order diagrams in $G$, are subdominant.

\begin{widetext}
Using the Feynman rules in the momentum space,
we obtain for single-loop contributions
%----------------------------------------
\bea\label{eq:feynmanrules_TT}
\Pi^{M}_{TT}(\omega_n)&=&
\frac{T}{4}\sum\limits_l\sum_{\alpha,\alpha'}\int\!
\frac{d\bm p}{(2\pi)^3}p_\alpha p_{\alpha'}
\Tr\bigl[\gamma_{\mu}
 G(\bm p, i\omega_l+i\omega_n)
\gamma_{\mu'} G(\bm p, i\omega_l)\bigr],\\
\label{eq:feynmanrules_TN}
\Pi^{M}_{TN}(\omega_n)&=&
\frac{T}{2}\sum\limits_l\sum_{\alpha}\int \!\frac{d\bm p}{(2\pi)^3}p_\alpha 
\Tr\bigl[\gamma_1 G(\bm p, i\omega_l+i\omega_n)
\gamma_{\mu} G(\bm p, i\omega_l)\bigr],\\
\label{eq:feynmanrules_JT}
\Pi^{M}_{JT}(\omega_n)&=&
\frac{T}{2}\sum\limits_l\sum_{\alpha}\int \!\frac{d\bm p}{(2\pi)^3}p_\alpha 
\Tr\bigl[\hat q\gamma_1 G(\bm p, i\omega_l+i\omega_n)
\gamma_{\mu} G(\bm p, i\omega_l)\bigr],\\
\label{eq:feynmanrules_NN}
\Pi^{M}_{NN}(\omega_n)&=&
T\sum\limits_l\int\! \frac{d\bm p}{(2\pi)^3}
\Tr\bigl[\gamma_1 G(\bm p, i\omega_l+i\omega_n)
\gamma_1 G(\bm p, i\omega_l)\bigr],\\
\label{eq:feynmanrules_JJ}
\Pi^{M}_{JJ}(\omega_n)&=&
T\sum\limits_l\int\! \frac{d\bm p}{(2\pi)^3}
\Tr\bigl[\hat q\gamma_1 G(\bm p, i\omega_l+i\omega_n)
\hat q\gamma_1 G(\bm p, i\omega_l)\bigr],\\
\label{eq:feynmanrules_JN}
\Pi^{M}_{JN}(\omega_n)&=&
T\sum\limits_l\int\! \frac{d\bm p}{(2\pi)^3} 
\Tr\bigl[\hat q\gamma_1 G(\bm p, i\omega_l+i\omega_n)
\gamma_1 G(\bm p, i\omega_l)\bigr],
\eea
% ----------------------------------------
\end{widetext}
where $\alpha, \alpha', \mu, \mu'$ assume values $1, 0$, with
$\mu\neq \alpha, \mu'\neq \alpha'$, \ie, the sum in
Eq.~\eqref{eq:feynmanrules_TT} contains four terms and those in Eqs.~\eqref{eq:feynmanrules_TN} and \eqref{eq:feynmanrules_JT} contain two terms, and $p_0=i\omega_l+i\omega_n/2$.
Here $G(\vecp, i\omega_l)$ is the dressed Matsubara Green's function
of quarks, and the summation goes over fermionic Matsubara frequencies
$\omega_l=\pi(2l+1)T-i\mu$, $l=0,\pm1,\ldots$ (note that we consider here isospin-symmetric matter, where both quarks have the same chemical potential $\mu$). The traces should be taken in Dirac, color, and flavor spaces.

The Matsubara summations appearing in these expressions can be cast into the general form
%----------------------------------------
\bea\label{eq:sums}
S_{\mu\nu}\left[f\right](\bm p,
i\omega_n)&=&T\sum\limits_l 
f(i\omega_l +i\omega_n/2)\nonumber\\
&&\hspace{-1.5cm}\times\Tr\bigr[\gamma_\mu G(\bm p, i\omega_l+i\omega_n)
\gamma_\nu G(\bm p, i\omega_l)\bigr],
\eea
%----------------------------------------
where $f(z)=z^k$ with $k=0,1,2$.
To perform the summation, we will use the spectral representation of the temperature Green's function
%----------------------------------------
\bea\label{eq:propagator}
G(\bm p, z)=\int_{-\infty}^{\infty}d\vep 
\frac{A(\bm p, \vep)}{z-\vep},
\eea 
%----------------------------------------
where $A(\bm p, \vep)$ is the spectral function of quarks; its finite
width, which effectively incorporates meson-mediated scattering,
allows dissipation to be accounted for through the one-loop diagram.
 After
performing a calculation of residues and an analytical continuation
($i\omega_n\to \omega +i\delta$), we find for Eq.~\eqref{eq:sums}
%---------------------------------------------------
\bea\label{eq:residue4}
S_{\mu\nu}[f](\bm p, \omega)&=&
\int_{-\infty}^{\infty} d\vep\int_{-\infty}^{\infty} d\vep'\, \Tr\bigl[\gamma_\mu A(\bm p, \vep')
\gamma_\nu A(\bm p, \vep)\bigr]
\nonumber\\
&\times&\frac{{\tilde n}
(\vep)f(\vep+\omega/2)-{\tilde n}(\vep')f(\vep'-\omega/2)}{\vep-\vep'+\omega+i\delta},
\eea
%---------------------------------------------------
where $n(\vep)=[e^{\beta (\vep-\mu)}+1]^{-1}$ is 
the Fermi distribution function, and ${\tilde n}(\vep)=n(\vep)-1/2$. 
Separating the real and imaginary parts in Eq.~\eqref{eq:residue4} via the Dirac identity and going to the limit $\omega\to 0$, we find
%--------------------------------------------------------- 
\bea\label{eq:difimsums}
\frac{d}{d\omega}{\rm Im} S_{\mu\nu}[f]
(\bm p, \omega)\bigg|_{\omega=0}=
\pi\int_{-\infty}^{\infty}d\vep\,
\frac{\partial n(\vep)}{\partial \vep } f(\vep)
{\cal T}_{\mu\nu}(\bm p, \vep),\nonumber\\
\eea
%--------------------------------------------------------- 
where 
%--------------------------------------------------------- 
\bea\label{eq:TraceA2}
{\cal T}_{\mu\nu}(\bm p, \vep)= \Tr\bigl[
\gamma_\mu A(\bm p, \vep)
\gamma_\nu A(\bm p, \vep)\bigr].
\eea
%---------------------------------------------------------

Substituting Eq.~\eqref{eq:sums}
into the correlation functions
\eqref{eq:feynmanrules_TT}--\eqref{eq:feynmanrules_JN}, we obtain
compact expressions in terms of Eq.~\eqref{eq:residue4}
%---------------------------------------------------
\bea\label{eq:feynmanrules1}
\Pi_{TT}(\omega)&=&
\frac{1}{4}\int\! \frac{d\bm p}{(2\pi)^3}\Big\{p_1^2
S_{00}[f=1]+2p_1S_{01}[f=\vep]\nonumber\\
&+&
S_{11}[f=\vep^2]\Big\},\\
\label{eq:feynmanrules2}
\Pi_{TN}(\omega)&=&
\frac{1}{2}\int\! \frac{d\bm p}{(2\pi)^3}
\Big\{p_1 S_{10}[f=1]+S_{11}[f=\vep]\Big\},\nonumber\\\\
\label{eq:feynmanrules3}
\Pi_{NN}(\omega)&=&
\int\! \frac{d\bm p}{(2\pi)^3}
S_{11}[f=1],\\
\label{eq:feynmanrules4}
\Pi_{JT}(\omega)&=&
\Pi_{TN}(\omega) 
\frac{\Tr\hat q}{N_f},\\
\label{eq:feynmanrules5}
\Pi_{JJ}(\omega)&=&
\Pi_{NN}(\omega)
\frac{\Tr\hat q^2}{N_f},\\
\label{eq:feynmanrules6}
\Pi_{JN}(\omega)&=&
\Pi_{NN}(\omega) 
\frac{\Tr\hat q}{N_f},
\eea
%--------------------------------------------------------- 
with $N_f$ being the number of flavors. Substituting Eq.~\eqref{eq:difimsums} into
Eqs.~\eqref{eq:feynmanrules1}--\eqref{eq:feynmanrules3} 
we obtain
%---------------------------------------------------
\bea\label{eq:feynmanrules7}
\frac{d}{d\omega}{\rm Im}\Pi_{TT}(\omega)\bigg|_{\omega=0}&=&
\frac{\pi}{4}\int_{-\infty}^{\infty}d\vep\,
\frac{\partial n(\vep)}{\partial \vep } \nonumber\\
&&\hspace{-2cm}\times\!\int\! \frac{d\bm
  p}{(2\pi)^3}\left(p_1^2
{\cal T}_{00}+2p_1\vep {\cal T}_{01}+
\vep^2 {\cal T}_{11}\right),\\
\label{eq:feynmanrules8}
\frac{d}{d\omega}{\rm Im}\Pi_{TN}(\omega)\bigg|_{\omega=0}&=&
\frac{\pi}{2}\int_{-\infty}^{\infty}d\vep\,
\frac{\partial n(\vep)}{\partial \vep } \nonumber\\
&&\hspace{-2cm}\times\!\int\! \frac{d\bm
  p}{(2\pi)^3}
\left(p_1 {\cal T}_{10}+\vep{\cal T}_{11}\right),\\
\label{eq:feynmanrules9}
\frac{d}{d\omega}{\rm Im}\Pi_{NN}(\omega)\bigg|_{\omega=0}&=&
\pi\int_{-\infty}^{\infty}d\vep\,
\frac{\partial n(\vep)}{\partial \vep } \!\int\! \frac{d\bm p}{(2\pi)^3}{\cal T}_{11}.\nonumber\\
\eea
%---------------------------------------------------------
Substituting Eqs.~\eqref{eq:feynmanrules4}--\eqref{eq:feynmanrules9} into Eqs.~\eqref{eq:corsigma_Q_m} and \eqref{eq:corkappa_m}
and then into \eqref{eq:sigma}--\eqref{eq:Qsigma},
we obtain the following expressions for
thermoelectric transport coefficients:
%--------------------------------------------------------- 
\bea\label{eq:sigma2}
\sigma &=&-\pi \frac{\Tr\hat q^2}{N_f}\int_{-\infty}^{\infty} 
d\vep\, \frac{\partial n(\vep)}{\partial \vep } \! \int\! \frac{d\bm p}
{(2\pi)^3}{\cal T}_{11}(\bm p,\vep),\\
\label{eq:sigmaQ2}
\sigma Q &=&-\frac{\pi}{2T}\frac{\Tr\hat q}{N_f}
\int_{-\infty}^{\infty}
d\vep\, \frac{\partial n(\vep)}{\partial \vep }\! \int\! \frac{d\bm p}
{(2\pi)^3}\Big[p_1{\cal T}_{10}(\bm p,\vep)\nonumber\\
&+&
(\vep -2h){\cal T}_{11}(\bm p,\vep)\Big],\\
\label{eq:kappa2}
\tilde{\kappa} &=&-\frac{\pi}{4T}\int_{-\infty}^{\infty}
d\vep\, \frac{\partial n(\vep)}{\partial \vep }\! \int\! \frac{d\bm p}{(2\pi)^3}\Big[p_1^2{\cal T}_{00}(\bm p,\vep)\nonumber\\
&+&
2p_1(\vep -2h){\cal T}_{01}(\bm p,\vep)+
(\vep -2h)^2 {\cal T}_{11}(\bm p,\vep)\Big].\quad
\eea
%--------------------------------------------------------- 
To compute the Dirac traces~\eqref{eq:TraceA2}, we use the 
general decomposition of the quark spectral function in terms of Lorentz components
%----------------------------------------------------------------
\bea\label{eq:spectral}
A(\bm p, p_0)=-\frac{1}{\pi}(mA_s+p_0\gamma_0 A_0-\bm p \cdot \bm\gamma A_v),
\eea
%----------------------------------------------------------------
where the coefficients $A_s=A_s(p,\vep)$, $A_0=A_0(p,\vep)$, $A_v=A_v(p,\vep)$ depend only on $\bm p^2$ and $\vep$ in isotropic medium. Note that the mass of the dressed quark $m$ is different from the current mass $m_0$ in the Lagrangian~\eqref{eq:lagrangian} because of the strong interactions~\cite{2005PhR...407..205B}.
%-------------------------------------------
Then we obtain for Eq.~\eqref{eq:TraceA2}
%-------------------------------------------
\bea\label{eq:traces}
\frac{\pi^2}{N_cN_f}{\cal T}_{\mu\nu}(\bm p, p_0) &=&
4(A_s^2m^2-A_0^2p_0^2+A_v^2\bm p^2)g_{\mu\nu}\nonumber\\
&+&8\Big[A_0^2p_0^2g_{\mu 0}g_{\nu 0} +A_v^2p_ip_jg_{\mu i}g_{\nu j}
\nonumber\\
&-&A_0A_vp_0p_i(g_{\mu i}
g_{\nu 0}+g_{\mu 0}g_{\nu i})\Big],\qquad
\eea
%-------------------------------------------
where we summed also over the quark flavor ($N_f$) and color ($N_c$)
numbers. Substituting these expressions into
Eqs.~\eqref{eq:sigma2}--\eqref{eq:kappa2} and performing the angular
integrations, we finally obtain
%----------------------------------------------------------------
\bea\label{eq:sigmafinal}
\sigma &=&\frac{10e^2 N_c}{27\pi^3}\int_{-\infty}^{\infty} \!
d\vep\, \frac{\partial n}{\partial \vep}\int_{0}^{\Lambda} dp\, p^2\nonumber\\
&\times&\Big[3A_s^2m^2 - 3A_0^2\vep ^2+A_v^2p^2\Big],\\
\label{eq:sigmaQfinal}
\sigma Q&=&-\frac{eN_c}{9\pi^3T}\int_{-\infty}^{\infty} \!
d\vep\, \frac{\partial n}{\partial \vep}\int_{0}^{\Lambda} dp\, p^2 \Big[2 A_0 A_v \vep p^2 \nonumber\\
&-& (\vep-2h)\left(3 A_s^2 m^2-3 A_0^2 \vep^2+A_v^2 p^2\right)\Big],\\
\label{eq:kappafinal}
\tilde{\kappa} &=&-\frac{N_cN_f}{6\pi^3T} \int_{-\infty}^{\infty}\!
d\vep\, \frac{\partial n}{\partial \vep}\int_{0}^{\Lambda} dp\, {p^2}\nonumber\\
&\times&\Big\{
2\big[A_0\vep +A_v(\vep -2h)\big]^2p^2 
\nonumber\\
&&
+\big(A_s^2m^2- A_0^2\vep^2+A_v^2p^2\big)
\left[p^2-3(\vep -2h)^2\right]\Big\},\quad\nonumber\\
\eea
%--------------------------------------------------------------
where we applied a 3-momentum cutoff, 
and took into account that $\Tr \hat{q}=e/3$ and $\Tr \hat{q}^2=5e^2/9$, see Eq.~\eqref{eq:charge}. 
Thus, the final computation of transport coefficients can be performed once the  three Lorentz components of the quark spectral function are known.

\section{Numerical results}
\label{sec:results}

Before presenting our key results, we briefly review the generic features of the two-flavor NJL phase diagram and the scattering processes relevant for the quark spectral functions. For further details, see Refs.~\cite{1961PhRv..124..246N,1961PhRv..122..345N,1991PrPNP..27..195V,1992RvMP...64..649K,2005PhR...407..205B,LKW15,Harutyunyan2017}.

\subsection{Quark spectral functions}

%--------------------------------------------------------
\begin{figure}[t] 
\begin{center}
\includegraphics[height=2cm]{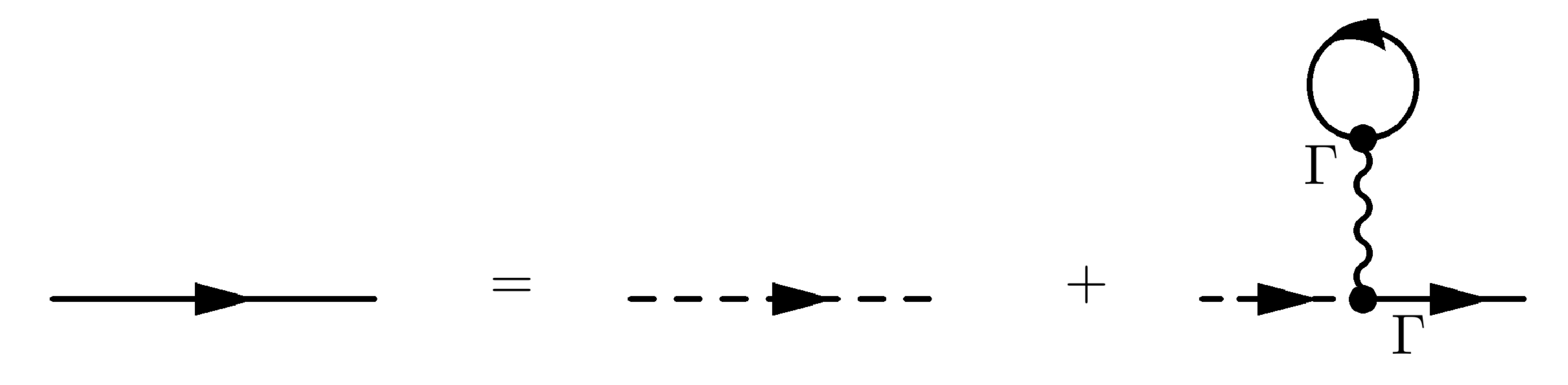}
\caption{The Dyson-Schwinger equation for the constituent quark 
mass. The dashed and solid lines stand for the bare and dressed 
quark propagators, respectively, and the vertex $\Gamma=1$. 
The wavy line represents the four-fermion coupling $G$.}
\label{fig:gap_eq} 
\end{center}
\end{figure}
%--------------------------------------------------------
%--------------------------------------------------------

The constituent (in-medium) quark mass at finite temperatures and chemical potentials is found from the Dyson-Schwinger equation
in the Hartree approximation, shown in Fig.~\ref{fig:gap_eq}
%---------------------------------------------------------------
\bea\label{eq:mass} 
m(T,\mu)=m_0-G\langle\bar{\psi}\psi\rangle,  
\eea
%---------------------------------------------------------------
with the quark condensate given by 
%---------------------------------------------------------------
\bea\label{eq:gap3}
\langle\bar{\psi}\psi\rangle= -\frac{mN_cN_f}{\pi^2}\int_0^\Lambda
dp\frac{p^2}{E_p}[1-n^+(E_p)-n^-(E_p)],~
\eea 
%---------------------------------------------------------------
where $E_p=\sqrt{p^2+m^2}$, and we redefined the Fermi distribution 
for quarks/antiquarks as $n^\pm(E_p)=[e^{\beta(E_p\mp\mu)}+1]^{-1}$.

\begin{figure*}[hbt] 
\begin{center}
\includegraphics[width=10.0cm,keepaspectratio]{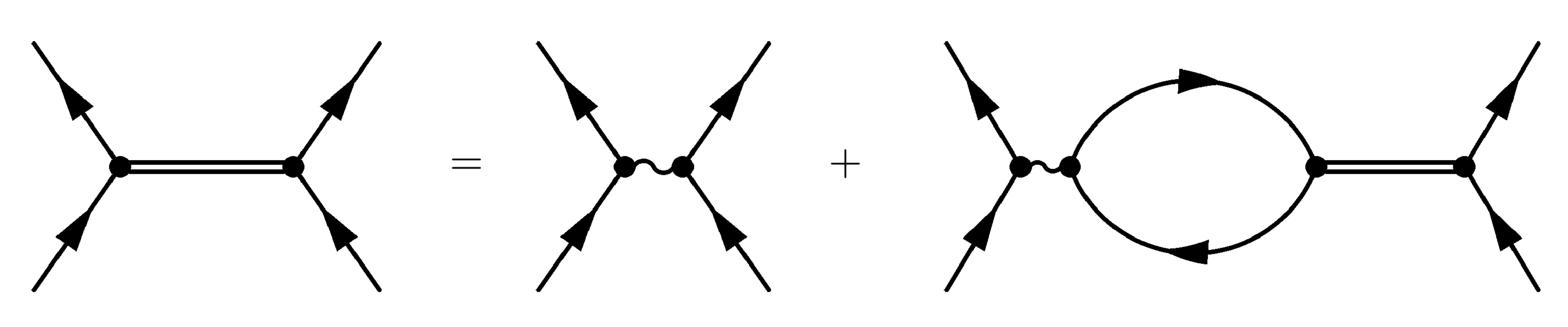}
\caption{The Bethe-Salpeter equation for mesonic modes. The double 
lines represent the dressed meson propagators, the solid lines stand 
for the dressed quark propagators, and the vertex assumes the values 
$\Gamma^0_{s}=1$ for $\sigma$ meson and $\Gamma^0_{ps}=i\bm\tau\gamma_5$
  for pions. }
\label{fig:BS_eq} 
\end{center}
\end{figure*}

 The dressed mesonic propagators are found from the Bethe-Salpeter 
 equation which resums a geometrical series  of quark-antiquark loops 
 diagrams, as shown in Fig.~\ref{fig:BS_eq}.
 The meson masses are then determined as the poles of the dressed 
meson propagators in real spacetime for vanishing momentum $\bm p=0.$
%--------------------------------------------------------
%--------------------------------------------------------

\begin{figure}[t] 
\begin{center}
\includegraphics[width=8cm,keepaspectratio]{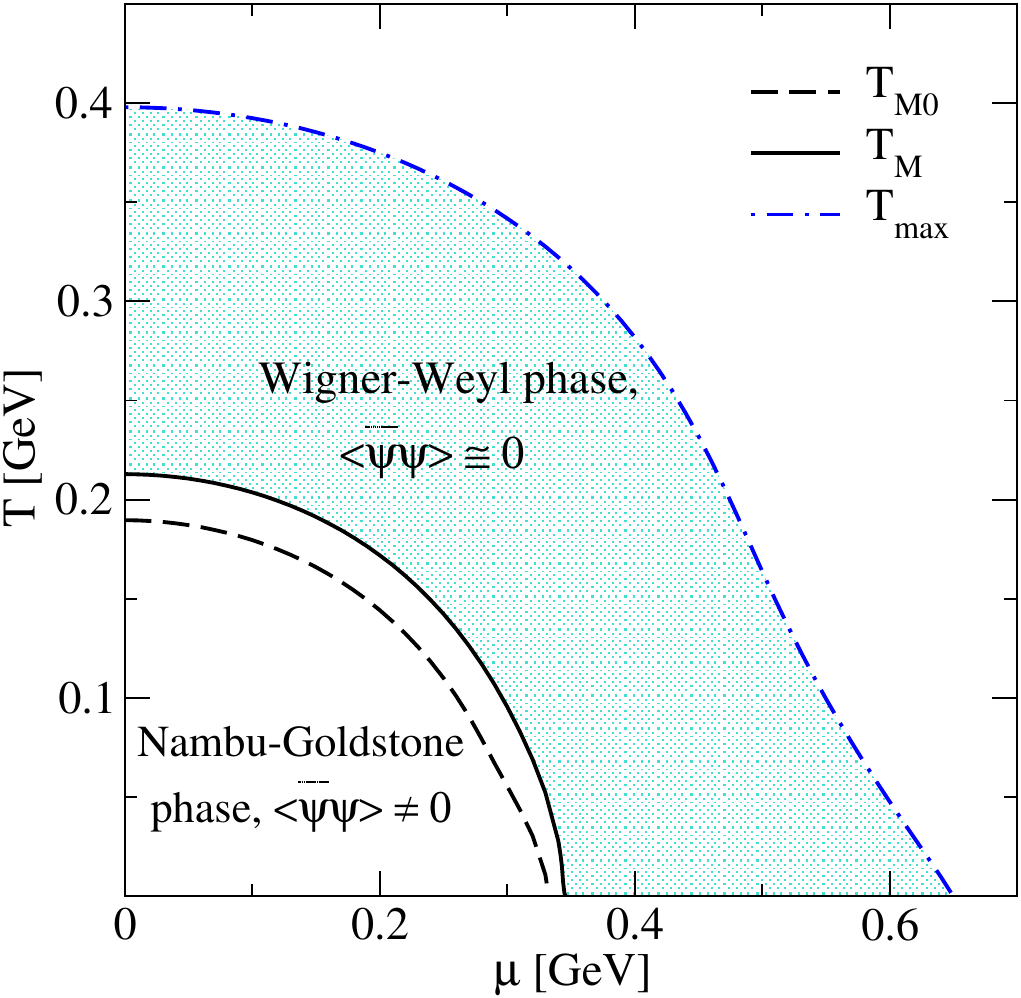}
\caption{ The phase diagram of strongly interacting quark matter at
  the leading order of $1/N_c$ expansion. The shaded area shows the region where our approximations are applicable. The inner bound
  is the Mott temperature $T_{\rm M}$ (in the case of broken chiral
  symmetry) or $T_{\rm M 0}\equiv T_c$ (in the case where chiral
  symmetry is intact), and the upper bound is $T_{\rm max}$ above
  which no mesonic modes are found. }
\label{fig:mott_temp} 
\end{center}
\end{figure}

Figure~\ref{fig:mott_temp} shows the phase diagram of the two-flavor
NJL model on the $\mu$-$T$ plane within the approximations used. The
shaded area separates the part of the phase diagram where the
dispersive effects of the meson decay into on-shell quark-antiquark
pair $\pi,\sigma\to q+\bar{q}$ are kinematically allowed. The inner
boundary of the shaded area is the so-called Mott temperature
$T_{\rm M}$ at which the pion mass equals twice the quark mass.  In
the case of exact chiral symmetry, i.e., $m_0 = 0$, the Mott
temperature coincides with the critical temperature $T_c$ of the
chiral phase transition. Above this temperature, we have
$\langle \bar{\psi} \psi \rangle = 0$ and $m = 0$, signaling the
restoration of chiral symmetry.  The upper boundary of the diagram is
the maximal temperature $T_{\rm max}$ above which mesonic modes do not
exist anymore within the zero-momentum pole approximation of the
${\cal O}(N_c)$ expansion, see Ref.~\cite{Harutyunyan2017} for
details. At $T = 0$ this line ends at the maximal value of the
chemical potential $\mu_{\rm max}=\Lambda$, where the meson masses are
$m_M = 2\Lambda$.
%--------------------------------------------------------------
%--------------------------------------------------------------

The three components of the quark spectral function $A_s$, $A_0$, and $A_v$
can be expressed in terms of the relevant components of the quark
self-energy according to the relations~\cite{LW14,LKW15,Harutyunyan2017}
%-------------------------------------------------------
\bea\label{eq:spectral_coeff}
A_i(p,p_0)=\frac{1}{d} \big[n_1\varrho_i 
-2n_2 (1+r_i) \big],\quad d=n_1^2+4n_2^2,
\eea
%-------------------------------------------------------
where $\varrho_i = {\rm Im}\Sigma_i$,
$r_i = {\rm Re}\Sigma_i$, $i=s,0,v$, and
%-------------------------------------------------------
\bea
\label{eq:N1}
n_1&=&p_0^2[(1+r_0)^2-\varrho_0^2]\nonumber\\
&-&\bm p^2[(1+r_v)^2-\varrho_v^2]
-m^2[(1+r_s)^2-\varrho_s^2],\\
\label{eq:N2}
n_2&=&p_0^2\varrho_0 (1+r_0) \nonumber\\
&-&\bm p^2\varrho_v(1+r_v)-m^2\varrho_s(1+r_s).
\eea
%--------------------------------------------------------

The three Lorentz components of the quark retarded/advanced 
self-energy are identified via (the asterisk denotes complex conjugation):
%-------------------------------------------------------
\bea\label{eq:selfenergy}
\Sigma^{R(A)} =m\Sigma_s^{(*)}
-p_0\gamma_0\Sigma_0^{(*)}
+\bm p\bm\gamma\Sigma_v^{(*)},
\eea 
%-------------------------------------------------------
and each of the Lorentz components is a sum of contributions 
by $\sigma$ and $\pi$ mesons as follows
%-----------------------------------------------------------------
\bea\label{eq:selfsum2} 
\Sigma_s= \Sigma^\sigma_s-3\Sigma^\pi_s,\quad
\Sigma_{0/v} =-\Sigma^\sigma_{0/v} -3\Sigma^\pi_{0/v}.
\eea
%-----------------------------------------------------------------

In this work, we will employ the quark spectral functions derived from
one-meson-exchange diagrams in the regime close to the chiral phase
transition line.  The imaginary parts of the on-shell quark and
antiquark self-energies which arise from the meson decay into
quark-antiquark pair were computed in
Refs.~\cite{LW14,LKW15,Harutyunyan2017}
%-------------------------------------------------------
\bea\label{eq:im_self}
\varrho_{j}^M(p)\Big\vert_{p_0=E_p}\! &=&
\frac{g^2_M}{16\pi p}
\int_{E_{\rm min}}^{E_{\rm max}}\! d E
\nonumber\\
&\times&\mathscr{T}_{j}
\left[n^-(E) + n_B(E+E_p)\right],\quad\\
\label{eq:im_self_anti}
\varrho^M_{j}(p)\Big\vert_{p_0=-E_p}\! &=&-
\frac{g^2_M}{16\pi p}
\int_{E_{\rm min}}^{E_{\rm max}}\! d E \nonumber\\
&\times&
\mathscr{T}_{j}\left[n^+(E) + n_B(E+E_p)\right],\quad
\eea
%-------------------------------------------------------
where $j=s,0,v$; the quark energy is defined as $E_p =\sqrt{p^2+m^2}$, 
$g_M$ is the in-medium quark-meson coupling with $M=\pi,\sigma$ 
(defined as the residue of the full meson propagator 
at vanishing momentum), $n_B(E)=(e^{\beta E}-1)^{-1}$ 
is the Bose distribution for mesons, and
%-------------------------------------------------------
\bea\label{eq:f_sv}
\mathscr{T}_{s}=1,\quad
\mathscr{T}_{v}=\frac{m_M^2-2m^2-2EE_p}{2p^2},\quad
\mathscr{T}_{0}=-\frac{E}{E_p}.\,
\eea
%-------------------------------------------------------
The integration limits in Eqs.~\eqref{eq:im_self} 
and \eqref{eq:im_self_anti} are  
%-------------------------------------------------------
\bea\label{eq:E_min_max}
 E_{{\rm min},{\rm max}}&=&
\frac{1}{2m^2}\left[(m_M^2-2m^2)  E_p \right. \nonumber \\
&  & \hspace*{1cm} \left. \pm pm_M\sqrt{m_M^2-4m^2}\right],
\eea  
%-------------------------------------------------------
where $m_M$ is the meson mass.  We note that
Eqs.~\eqref{eq:im_self}–\eqref{eq:E_min_max} are valid only above the
Mott (critical) temperature, where the condition $m_M \ge 2m$\ is
satisfied. In the on-shell approximation, the full quark–antiquark
self-energy can now be expressed as
%-------------------------------------------------------
\bea\label{eq:im_self_onshell}
\varrho_j(p_0,p) =
\theta (p_0)\varrho^+_j(p)+
\theta (-p_0)\varrho^-_j(p),
\eea
%-------------------------------------------------------
with $\varrho^\pm_j(p)=\varrho_j(p_0=\pm E_p,p)$, where each of 
$\varrho_j$ is summed over the mesons according to Eq.~\eqref{eq:selfsum2}.
From Eqs.~\eqref{eq:im_self} and \eqref{eq:im_self_anti} we find
$\varrho^-_j(\mu, p)=-\varrho^+_j(-\mu, p)$, therefore
%-------------------------------------------------------
\bea\label{eq:relation_rho}
\varrho_j(\mu, -p_0, p) = -\varrho_j(-\mu, p_0, p).
\eea
%-------------------------------------------------------
The real parts $r_i$ of the  self-energy are of next-to-leading order in  ${\cal O} (N_c^{-1})$ power counting scheme and can be dropped.

%-------------------------------------------------
\begin{figure}[t] 
\begin{center}
\includegraphics[width=8.0cm,keepaspectratio]{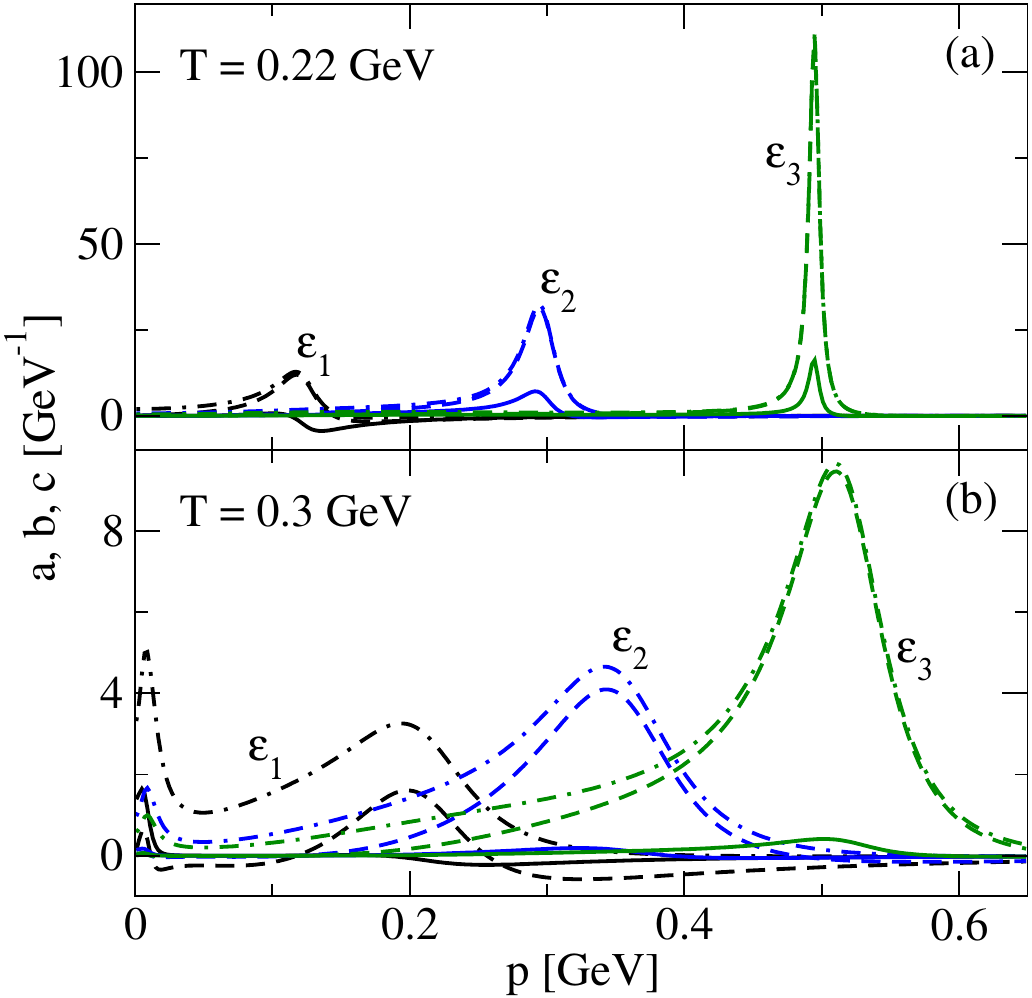}
\caption{ Dependence of three Lorentz components of the quark spectral functions $a=-mA_s$ (solid lines), $b=-\varepsilon A_0$ (dash-dotted lines), and $c=-pA_v$ (dashed lines) on the quark momentum at $\mu=0$ for $T=0.22$~GeV (a) and $T=0.3$~GeV (b). These spectral functions are shown at three energies
$\varepsilon_1 = 0.1$, $\varepsilon_2 = 0.3$, and
$\varepsilon_3 = 0.5$ GeV, as indicated in the plot. }
\label{fig:spectral} 
\end{center}
\end{figure}
%------------------------------------------------------------------
We also observe that the temporal and vector components of the spectral function are very close numerically at high energies, while the scalar component remains consistently smaller. This suggests that the primary contribution to the transport coefficients will come from
the temporal and vector components of the spectral functions.

The three Lorentz components of the spectral function obtained in Refs.~\cite{LKW15, Harutyunyan2017} are shown in Fig.~\ref{fig:spectral} as functions of the quark momentum for
three values of the quark (off-shell) energy: $\varepsilon_1 = 0.1$,
$\varepsilon_2 = 0.3$, and $\varepsilon_3 = 0.5$ GeV. The chemical potential is fixed at $\mu=0$; however, the general features of the spectral functions also remain the same for finite values of $\mu$. We see that the
spectral functions are peaked around
$p\simeq \varepsilon$, but the peaks are broadened with increasing temperature as more phase space is opened for the quark-meson scattering with increasing $T$.

\begin{table*}[t]
\centering
\caption{Comparison of thermopower $Q$ obtained in different theoretical approaches. 
The quoted values correspond to typical magnitudes in the indicated temperature and 
chemical-potential ranges. For $Q(T)$ and $Q(\mu)$, “Increasing” (respectively, “Decreasing”) indicates that the thermopower grows (respectively, decreases) with temperature $T$ or chemical potential $\mu$. Center dots $\dots$ indicate that no clear trend is reported in the reference.}
\label{tab:thermopower_comparison}
\begin{tabular}{lccccccc}
\hline\hline
Ref. 
& Model / approach 
& d.o.f.
& $T$ range (GeV) 
& $\mu$ range (GeV) 
& $Q$ 
& $Q(T)$ 
& $Q(\mu)$ \\
\hline
%=====================================
This work 
& NJL/Kubo 
& 2$f$ quarks + mesons 
& $0.17$--$0.30$ 
& $0.05$--$0.20$ 
& ($-$) $5$--$45$
& Increasing 
& Decreasing \\
\hline
%=====================================
\cite{Dey2020}
& pQCD / RTA
& 3$f$ quarks + gluons
& $0.17$--$0.40$
& $0.03$--$0.05$
& (+) $0.01$--$0.08$
& Decreasing
& Increasing \\ 
%=====================================
\cite{Kurian2021}
& pQCD / RTA
& 3$f$ quarks + gluons 
& $0.26$--$0.40$
& $0.08$--$0.10$
& ($-$) $5$--$35$
& Decreasing
& Increasing \\
%=====================================
\hline
\cite{Rath2025}
& pQCD/ RTA
& 3$f$  quarks + gluons 
& $0.17$--$0.64$
& $0.04$--$0.08$
& (+)  $0.01$--$0.10$
& Decreasing
& Increasing \\ 
%=====================================
\cite{Shaikh2025}
& pQCD / RTA
& 3$f$ quarks + gluons
& $0.17$--$0.65$
& $0.04$--$0.08$
& (+) $0.01$--$16$
& Decreasing
& Increasing \\
\hline
%=====================================
\cite{Khan2024}
& pQCD / RTA
& 3$f$ quarks + gluons
& $0.17$--$0.50$
& $0.04$
& (+) $0.35$--$1$
& Decreasing
& Increasing \\
\hline
%=====================================
\cite{Singh2025}
& pQCD / RTA
& 2$f$ quarks + mesons
& 0.16--0.32
& 0.10--0.60
& ($-$) $0.1$--$5$
& Increasing
& Increasing \\
\hline
%=====================================
\cite{Abhishek2020}
& NJL / RTA
& 2$f$ quarks + mesons
& 0.10--0.26
& 0.10--0.20
& ($-$) $2$--$20$
& Decreasing
& Increasing \\
\hline
%=====================================
\cite{Zhang2022}
& NJL / RTA
& 2$f$ quarks + mesons
& 0.10--0.40
& 0.10 
& (+) 7
& Const. 
& $\dots$\\
\hline\hline
\end{tabular}
\end{table*}

\subsection{Thermoelectric coefficients}

To evaluate thermoelectric coefficients, we need the enthalpy per particle $h=\mu+Ts/n$, which
within the $1/N_c$ (quasiparticle) approximation is given by the
formulas~\cite{2005PhR...407..205B,Harutyunyan2017}
%------------------------------------------------------------
\bea\label{eq:enthalpy}
h =\frac{N_cN_f}{\pi^2 n}\int_0^\infty p^2dp\left(E_p+\frac{p^2}{3E_p}\right)
\left[n^+(E_p)+n^-(E_p)\right],\nonumber\\
\eea
%------------------------------------------------------------ 
where $E_p=\sqrt{p^2+m^2}$, and
%------------------------------------------------------------
\bea\label{eq:quark_number_density}
n = \frac{N_cN_f}{\pi^2}\int_0^\infty 
p^2dp\left[n^+(E_p)-n^-(E_p)\right]
\eea
%------------------------------------------------------------
is the net quark number density. Note that the enthalpy per particle diverges in the limit where chemical potential tends to zero, as this implies vanishing quark number density. 

\subsubsection{Thermopower}

Let us now turn to the discussion of numerical results of thermopower
or, equivalently, the Seebeck coefficient, given by
formulas~\eqref{eq:sigmafinal} and \eqref{eq:sigmaQfinal}. First, we
consider the general structure of
Eqs.~\eqref{eq:sigmafinal}--\eqref{eq:kappafinal}. For the given
$\vep$, the expressions under the momentum integral display an
oscillatory structure, peaking around $p\simeq \vep$, as follows from
the form of the spectral functions. The height of peaks rapidly
increases with increasing $\vep$. As a consequence, the inner
(momentum) integrals in expressions
\eqref{eq:sigmafinal}--\eqref{eq:kappafinal} are rapidly increasing
functions of $|\vep|$ as long as $\vert\vep\vert\leq\Lambda$. For
energies larger than $\Lambda$, the peaks are outside the integration
limits (due to momentum cutoff), and the integral sharply decreases
with $\vep$. The outer integration contains the factor
$\partial n(\vep)/\partial\vep$ which has a broad, bell-shaped form
around its maximum value at $\vep = \mu$.

The dependence of the conductivities $\sigma$ and $\tilde{\kappa}$ on $T$ and $\mu$ was plotted and discussed in detail in Ref.~\cite{Harutyunyan2017}. These quantities were shown to decrease with increasing temperature for arbitrary values of the chemical potential, as a result of the broadening of the spectral functions. The same argument also holds for Eq.~\eqref{eq:sigmaQfinal}. In particular, they all diverge in the vicinity of the Mott temperature, because in the limit $T\to T_M$ the decay cross section of the mesons tends to zero, as a result of which the spectral functions transform into $\delta$-functions and lead to infrared divergence in Eqs.~\eqref{eq:sigmafinal}--\eqref{eq:kappafinal}.

%-------------------------------------------------
\begin{figure}[t] 
\begin{center}
\includegraphics[width=8cm,keepaspectratio]{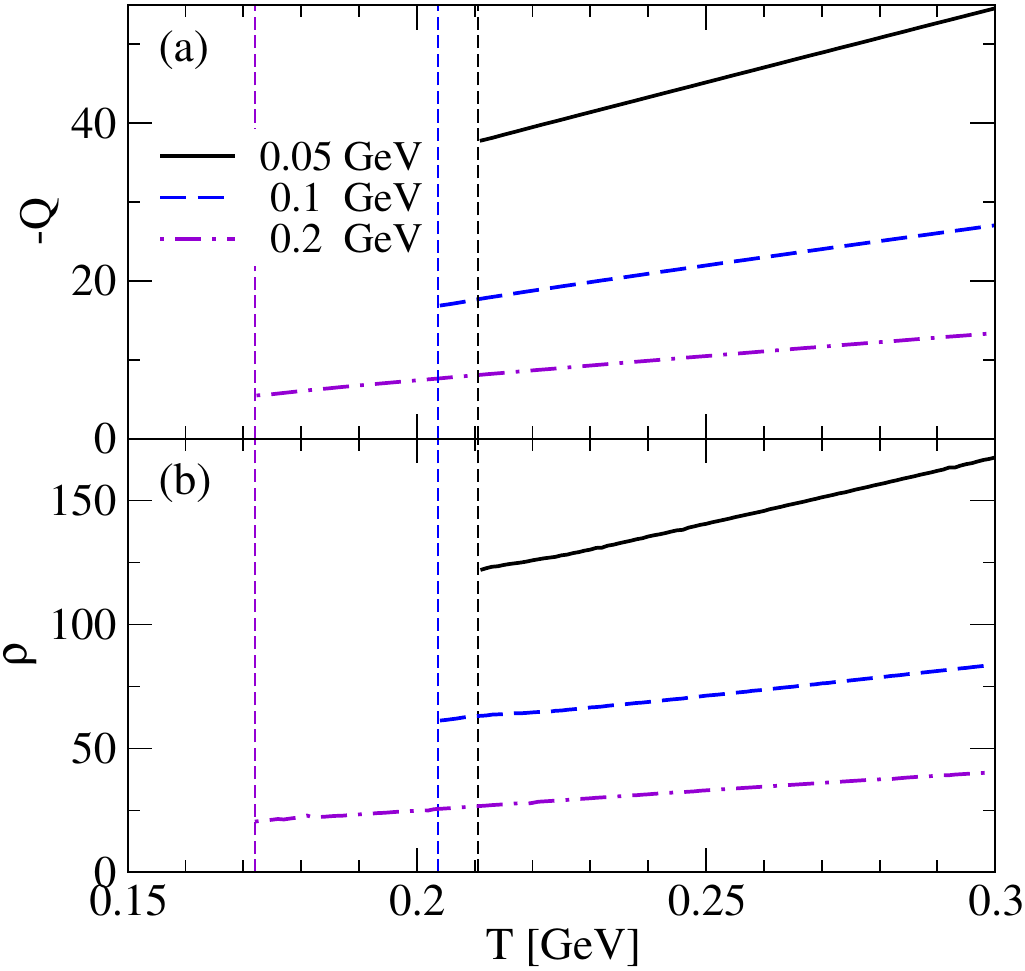}
\caption{ Thermopower (a) and the Thomson coefficient (b) as functions of the temperature at various values of the chemical potential  indicated on the plots. The  vertical lines show the Mott temperature at the given value of $\mu$. }
\label{fig:Q_temp} 
\end{center}
\end{figure}
%-------------------------------------------------
%-------------------------------------------------
\begin{figure}[!]
\begin{center}
%\vspace{0.6cm}
\includegraphics[width=7.75cm,keepaspectratio]{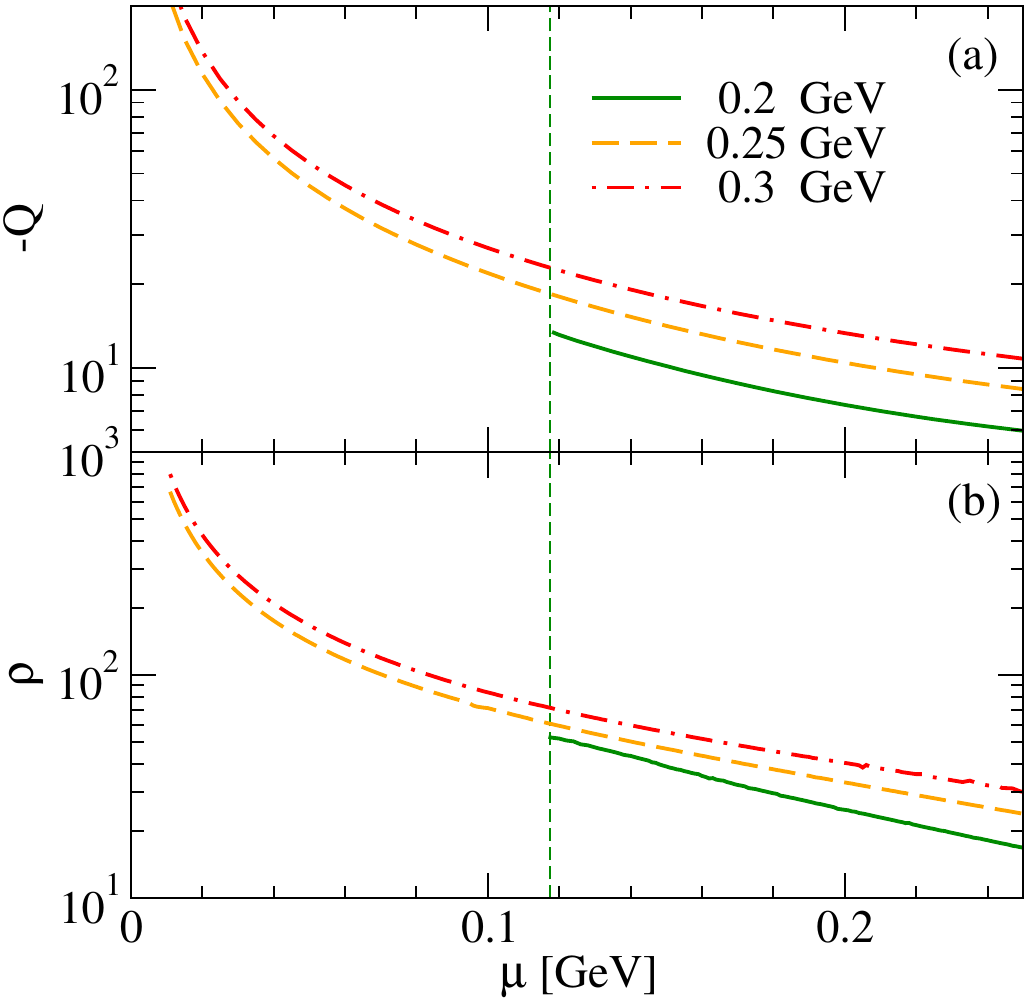}
\caption{ The dependence of (a) thermopower  and (b) Thomson coefficient
on the chemical potential at various temperatures indicated on the plots. The vertical line shows the value of the chemical potential on the Mott line for temperature $T=0.2$~GeV. %where the temperature approaches the Mott temperature.  
}
\label{fig:Q_mu} 
\end{center}
\end{figure}
%-------------------------------------------------

Figure \ref{fig:Q_temp} (a) shows the temperature dependence of the modulus of thermopower $-Q$ for constant values of the chemical potential. As $Q$ is given by the ratio of two correlation functions \eqref{eq:sigmaQfinal} and \eqref{eq:sigmafinal}, their dependencies on the spectral functions, as well as the divergencies in the vicinity of the Mott temperature, are almost completely canceled.  As a result, $Q$ depends weakly on the microscopic interactions and has thermodynamic nature. 
We find that $-Q$ increases almost linearly with increasing temperature. Approximating the integrands in formulas \eqref{eq:sigmaQfinal} and \eqref{eq:sigmafinal} as $pA_v\simeq \vep A_0\gg mA_s$, we obtain the following simple approximation:
%------------------------------------------
\bea\label{eq:Q_estimate}
Q\simeq \frac{3(\bar{\vep}-h)}{5eT}\simeq -\frac{7\pi^2 T}{25e\mu},
\eea 
%-------------------------------------------
where $\bar{\vep}$ is the average characteristic quark energy. The second approximation is applicable in the case where $T\gtrsim\mu$,
where $h\simeq 7\pi^2 T^2/15\mu\ge
\bar{\vep}$~\cite{Harutyunyan2017}. In this regime, the absolute value of the thermopower increases linearly with temperature, in good
agreement with the results shown in Fig.~\ref{fig:Q_temp}. From
Eq.~\eqref{eq:Q_estimate}, we also see that $Q$ is negative at small
to moderate values of $\mu$, but becomes positive at larger values of
$\mu$, which are not considered here.

Table~\ref{tab:thermopower_comparison} provides a comparison
  of the general trends and quantitative comparison of the thermopower
  obtained in different theoretical
  approaches~\cite{Dey2020,Kurian2021,Rath2025,Shaikh2025,Khan2024,Singh2025,Abhishek2020,Zhang2022}.
  Our results, computed within the NJL model above the Mott
  temperature, show relatively large negative values of $Q$ in the
  range $(-5)$--$(-45)$, with a clear trend of increasing magnitude
  with temperature and decreasing magnitude with chemical
  potential. This behavior contrasts with the perturbative quantum
  chromodynamics (pQCD) combined with Boltzmann-equation-based
  relaxation-time approximation (RTA) results reported in
  Refs.~\cite{Dey2020,Singh2025,Shaikh2025,Khan2024,Kurian2021}, where
  the thermopower is generally smaller in magnitude, often positive,
  and exhibits the opposite qualitative trend with respect to $\mu$,
  i.e., $Q$ increases with chemical potential. In particular, for
  chemical potentials $\mu \leq 100$~MeV, our results for the absolute
  value of the Seebeck coefficient exceed those reported in
  Refs.~\cite{Dey2020,Rath2025,Shaikh2025,Khan2024}, and
  \cite{Singh2025}  by approximately 4,
  3, 1, 2, and 2 orders of magnitude, respectively.  Similarly, the
  Seebeck coefficients obtained in other NJL-based works
  \cite{Abhishek2020,Zhang2022} show trends comparable to ours,
  confirming the robustness of the RTA-NJL approach in the
  high-temperature, quark-meson phase of the model.  The differences
  in the absolute value of $Q$ with pQCD/RTA calculations can be
  attributed to the inclusion of gluon degrees of freedom, the use of
  energy-dependent relaxation times, and the different microscopic
  scattering mechanisms underlying these approaches. Overall, our
  results highlight that the NJL framework predicts substantially
  larger thermoelectric response in the quark-meson plasma than
  pQCD-based models.   It is worth noting that the increase
  of $Q$ with temperature observed in our model contrasts with the
  results of Refs.~\cite{Dey2020,Rath2025,Singh2025,Shaikh2025,Khan2024} based on pQCD/RTA approach.  The difference arises from the fact that the
  quasiparticle widths are parametrically small and scattering is
  dominated by leading-order perturbative processes, which tend to
  suppress thermoelectric response. In contrast, our approach
  incorporates nonperturbative effects through finite spectral widths
  of the quark propagators, leading to enhanced interaction rates and,
  consequently, larger values of the Seebeck coefficient.

 Turning to the comparison with the NJL/RTA approaches, we note that 
 the qualitative dependence of the
 thermopower found in this work on temperature and chemical potential is
 similar to that reported in Ref.~\cite{Kurian2021};
 however, the absolute magnitude of our results is larger by about a
 factor of 4, which can be traced to the different treatment of
 quasiparticle damping and medium effects.

Furthermore,  we find very good agreement, both qualitative and
quantitative, between our results for the thermopower and those of
Ref.~\cite{Abhishek2020}, which employed the same NJL model with
quark–meson scattering above the Mott temperature, but used Boltzmann
theory for weakly coupled quasiparticles to compute the transport
coefficients. Nevertheless, the results of~\cite{Abhishek2020} for the electrical and thermal conductivities differ significantly from our findings discussed in
Ref.~\cite{Harutyunyan2017}. The agreement between the thermopower
$Q$ in weakly and strongly interacting regimes, despite the strong
discrepancy in conductivities, can be attributed to the fact that 
the thermopower is a ratio of two conductivities. As a result, 
the influence of the strength of microscopic  interactions on $Q$ is
relatively weak. We also note that the thermoelectric contribution to
the thermal conductivity, $\tilde{\kappa} - \kappa = T Q^2 \sigma$,
which was neglected in Ref.~\cite{Harutyunyan2017}, provides only a
small correction of the order of 10\%–15\%.

\subsubsection{The Thomson coefficient}

Next, we turn to the discussion of the Thomson coefficient $\rho$, defined in Eq.~\eqref{eq:Thomson}. Since it is numerically easier to obtain $Q$ as a function of $T$ and $\mu$, we compute the Thomson coefficient using the chain rule:
%------------------------------------------------------------
\bea\label{eq:Thomson1}
\rho&=&-T\left(\frac{\partial Q}{\partial T}\right)_n\nonumber\\
&=&\beta\left(\frac{\partial Q}{\partial \beta}\right)_\mu+\beta\left(\frac{\partial Q}{\partial \mu}\right)_\beta\left(\frac{\partial \mu}{\partial \beta}\right)_n,
\eea
%------------------------------------------------------------
where the derivative $({\partial \mu}/{\partial \beta})_n$ can be computed from Eq.~\eqref{eq:quark_number_density}~\cite{Harutyunyan2017}
%-------------------------------------------------
\bea
\beta\left(\frac{\partial \mu}{\partial \beta}\right)_n=
\frac{\int_{0}^{\infty}p^2dp\, \big[(E_p-\mu)n^+\tilde
n^+-(E_p+\mu)n^-\tilde n^-\big]}{\int_{0}^{\infty}p^2dp\,
\big[n^+\tilde n^++ n^-\tilde n^-\big]},\hspace{-0.5cm}\nonumber\\
\eea
% -------------------------------------------------
where $\tilde n^{\pm} \equiv 1-n^{\pm}$.
Numerically, we find that the first term in Eq.~\eqref{eq:Thomson1} accounts for about one-third of the total $\rho$. As $Q\propto T$ for small chemical potentials, then the first term in Eq.~\eqref{eq:Thomson1} is almost equal to $-Q$, and we obtain a rough estimate $\rho\simeq -3Q$.
Consequently, the Thomson effect becomes stronger at higher temperatures: for the same current and temperature gradient, more heat is released or absorbed. Note also that the Thomson coefficient is always positive in the regime of interest. This result contrasts with that of Ref.~\cite{Singh2025}, where the Thomson coefficient may be either positive or negative, depending on the value of the chemical potential. Moreover, the magnitude of the Thomson
coefficient obtained in our work, $|\rho|$, exceeds the values reported in Ref.~\cite{Singh2025} by 1--2 to orders of magnitude.

The dependence of thermopower and the Thomson coefficient on the chemical potential for fixed values of the temperature is shown in Fig.~\ref{fig:Q_mu}. Both coefficients decrease with the chemical potential and tend to infinity in the limit $\mu\to 0$. The observed divergence at vanishing chemical potential is similar to the behavior of the thermal conductivity~\cite{Harutyunyan2017} and arises due to the terms containing $h$ in expressions \eqref{eq:sigmaQfinal} and \eqref{eq:kappafinal}. Within the domain of temperatures and chemical potentials of investigation, the minimum value of enthalpy per particle is nearly $h_{\rm min}\simeq 0.8$ GeV~\cite{Harutyunyan2017}, which exceeds the cutoff parameter $\Lambda\simeq 0.65$ GeV, which serves as a characteristic energy scale for quarks. Thus, it can be concluded that the dominant terms in $Q$ and $\rho$ are those containing $h$. The physical reason for this divergence is the disappearance of the rest frame of the particles in the $n\to 0$ limit, relative to which the thermal current is defined~\cite{1985PhRvD..31...53D}.

%-------------------------------------------------
\begin{figure}[t] 
\begin{center}
\includegraphics[width=8cm,keepaspectratio]{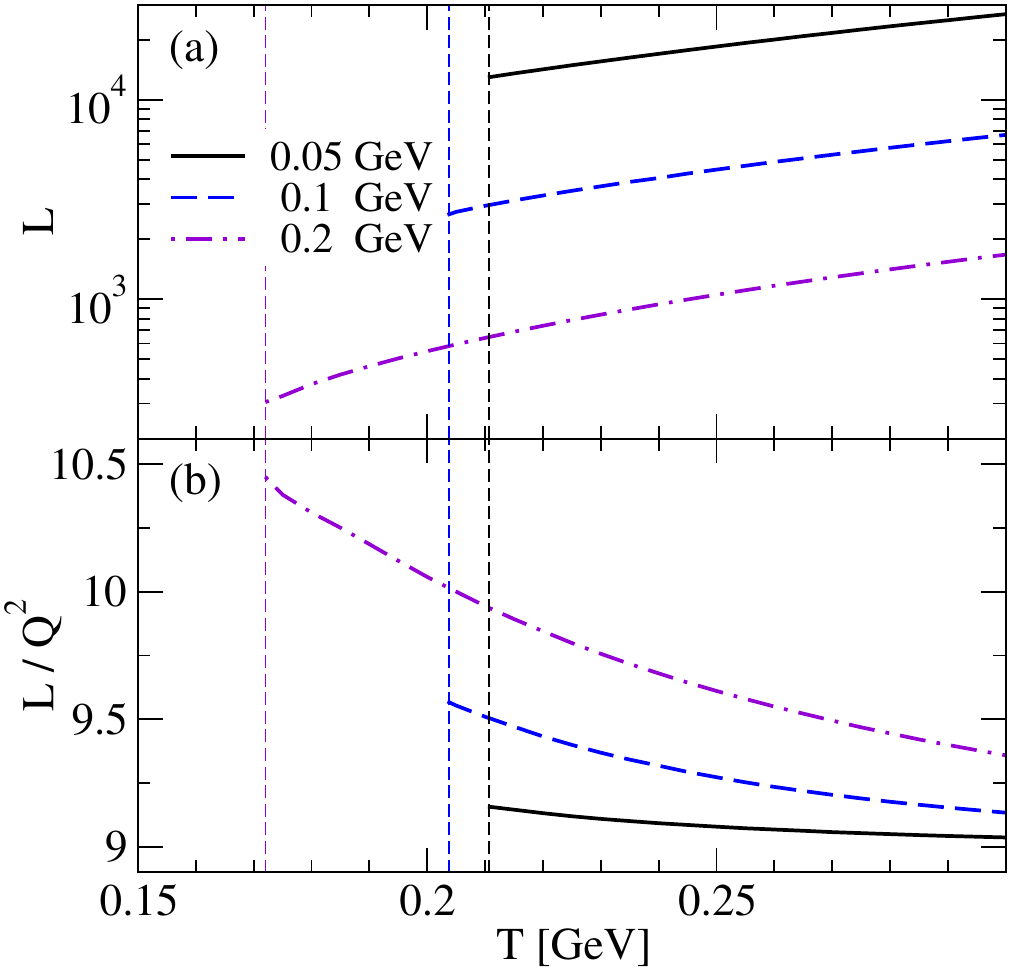}
\caption{ The Lorenz ratio $L=\kappa/\sigma T$ (a) and the ratio $L/Q^2$ (b) as functions of the temperature at various values of the chemical potential indicated on the plots. The  vertical lines show the Mott temperature at the given value of $\mu$. }
\label{fig:L_temp} 
\end{center}
\end{figure}
%-------------------------------------------------
\begin{figure}[t] 
\begin{center}
\includegraphics[width=8cm,keepaspectratio]{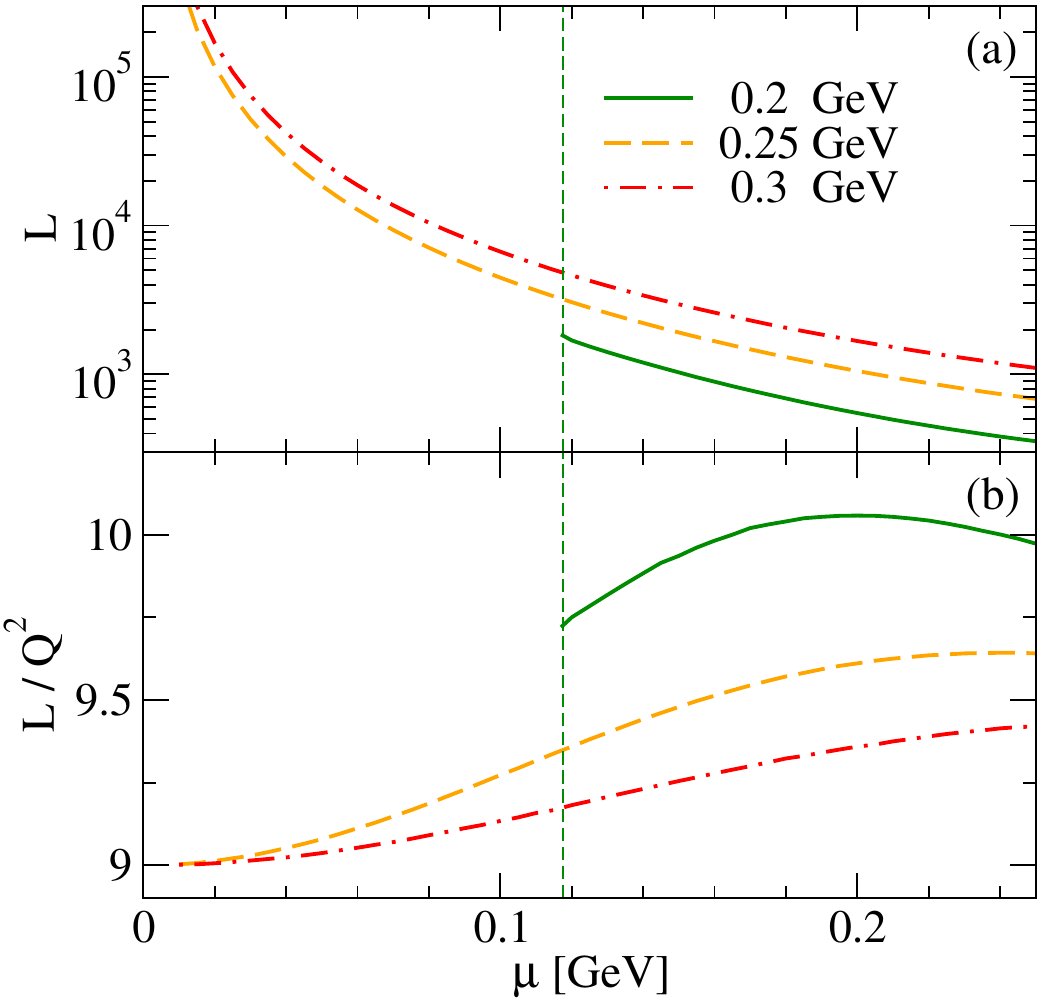}
\caption{ The Lorenz ratio $L=\kappa/\sigma T$(a) and the ratio $L/Q^2$ (b) as functions of the chemical potential at various values of the temperature indicated on the plots. The vertical line shows the value of the chemical potential on the Mott line for temperature $T=0.2$~GeV. }
\label{fig:L_mu} 
\end{center}
\end{figure}
%-------------------------------------------------

\subsubsection{The Lorenz ratio}

As shown earlier in Ref.~\cite{Harutyunyan2017}, the
  Wiedemann–Franz law for the Lorenz ratio $L = \kappa / (\sigma T)$
  does not hold in hot relativistic quark matter in the regime
  considered in this work. This violation originates from the
  divergence of the thermal current at small chemical potentials. A
  simple estimate of the Lorenz ratio was also provided in
  Ref.~\cite{Harutyunyan2017}
%------------------------------------------
\bea\label{eq:L_estimate}
L\simeq \frac{N_f}{\Tr\hat q^2} \frac{h^2}{T^2}\simeq \frac{98\pi^4}{125e^2}\frac{T^2}{\mu^2}.
\eea 
%-------------------------------------------
Although this estimate was derived for the regime $\mu\ll T$, the
exact numerical results suggest that it is valid also for larger
domain of chemical potentials $\mu\lesssim T$.  This can be seen from
Figs.~\ref{fig:L_temp}(a) and \ref{fig:L_mu}(a), which show
the exact Lorenz ratio as a function of temperature and chemical
potential, respectively.

We recall that for classical nonrelativistic electron-ion plasma the Lorenz ratio has the constant value $L_{\rm WF}=3/2e^2$, according to the 
Wiedemann-Franz law. However, in a multicomponent systems such as QGP, the
Lorenz ratio should be modified by substitution $e^2\to \Tr\hat q^2/N_f$ to account for different vales of electric charges of different quark species in the plasma~\cite{Rai2025}. Thus, the classical Lorenz ratio for two-flavor quark matter would be $L_{\rm WF}=27/5e^2\simeq 59$ (recall that in the units we use $e^2=4\pi\alpha$).
From Figs.~\ref{fig:L_temp} and \ref{fig:L_mu} we see that at $\mu\ll T$, the Lorenz ratio $L\gg L_{\rm WF}$, as found also in~\cite{Rai2025}. However, the Wiedemann-Franz law may be restored in the regime of degenerate quark matter with $\mu\gg T$, as mentioned in Refs.~\cite{Lucas2018,Rai2025} (in this regime for two-flavor quark matter one obtains $L_{\rm WF}=6\pi^2/5e^2\simeq 129$, instead of its value $L_{\rm WF}=\pi^2/3e^2$ for ordinary metals).

Comparing Eqs.~\eqref{eq:Q_estimate} and \eqref{eq:L_estimate}, we find a simple approximate relation $L /Q^2\simeq 10$. Figures~\ref{fig:L_temp}(b) and \ref{fig:L_mu}(b) show the exact value of $L/Q^2$ as a function of temperature and chemical potential, respectively. We see that, in the regime where $\mu\ll T$, this ratio tends to a constant value $L /Q^2\to 9$, and its variations are small in the whole range of temperatures and chemical potentials considered here (note that in degenerate gas $L /Q^2\to \infty$ as $T\to 0$, because
$Q\propto T$ in that regime~\cite{Ziman1979}).

\subsubsection{Generated thermoelectric fields}

Using the numerical results obtained above, we can make rough estimates of electric fields which will be generated in quark matter due to temperature gradients. A simple estimate gives $E\simeq T|Q|/l$, where $l$ is the characteristic average length scale of thermal gradients. Substituting here Eq.~\eqref{eq:Q_estimate}, we obtain
%------------------------------------------
\bea\label{eq:E_generated}
eE \simeq 
\frac{3T^2}{\mu l},
\eea 
%-------------------------------------------
where in the second step we assumed a value $l\simeq 10$~fm typical for heavy-ion collisions. For typical values $0.2 \leq T\leq 0.3$
and $0.1 \leq \mu\leq 0.2$~GeV, we find
$1.2\cdot 10^{-2}\lesssim eE \lesssim 5.4\cdot 10^{-2}$~GeV$^2$, or, in more relevant $m_\pi^2\simeq 2\cdot 10^{-2}$~GeV$^2$ units
%------------------------------------------
\bea\label{eq:E_generated1}
0.6 \lesssim \frac{eE}{m_\pi^2} \lesssim 2.7.
\eea 
%-------------------------------------------
This estimate is close to the values of electric fields induced by thermoelectric effects reported in Ref.~\cite{Singh2024}, where the hydrodynamic equations for an expanding QGP fireball created in head-on heavy-ion collisions were solved for both Bjorken and Gubser flow scenarios. These estimated electric fields underscore the importance of including thermoelectric effects in the phenomenology of heavy-ion collisions.

\section{Summary}
\label{sec:summary}

In this work, we studied the thermoelectric power and the Thomson
coefficient of quark matter within the two-flavor NJL model using the
Kubo formalism. We have derived the relevant Kubo formulas for
thermoelectric coefficients  starting from the quantum Liouville
equation for the density matrix, assuming the simultaneous presence of a weak external electric field and thermodynamic gradients. The relevant
correlation functions are computed for a relativistic quark plasma with
full quark propagator and within the $1/N_c$ approximation to the multiloop contributions, which implies that the leading-order contributions to the correlation functions arise from single-loop diagrams. The full Lorentz structure of quark spectral functions is taken into account. Here we used quark spectral functions derived from one-meson-exchange processes above the Mott transition temperatures~\cite{LKW15,Harutyunyan2017}.
Numerical results for the thermopower and the Thomson coefficient were obtained over a range of temperatures and chemical potentials relevant primarily to heavy-ion collisions. The results show that both coefficients increase nearly linearly with temperature, indicating that higher temperatures enhance the system’s ability to generate electric fields in response to a thermal gradient. We also find that these coefficients diverge in the limit of vanishing chemical potential. This behavior is a direct consequence of the fact that the particle rest frame, relative to which heat transport is defined, becomes ill defined in this limit.

We further estimated the magnitude of the electric fields generated in quark matter by the thermoelectric (Seebeck) effect and find that the effect is more pronounced at higher temperatures and relatively low chemical potentials. Finally, we compared our results with previous calculations performed within the RTA. We find that the Seebeck coefficient obtained here is larger than perturbative QCD estimates, while remaining consistent in order of magnitude with earlier NJL-model studies that employed again in the RTA.

\section*{Acknowledgements}

The authors acknowledge support through the collaborative research Grant
No. 24RL-1C010 provided by the Higher Education and Science Committee
(HESC) of the Republic of Armenia through the “Remote Laboratory”
program. A.~S. also acknowledges support from the Deutsche
Forschungsgemeinschaft Grant No. SE 1836/6-1 and the Polish National
Science Centre (NCN) Grant No. 2023/51/B/ST9/02798.

\appendix

\section{Solving the Liouville equation}
\label{app:Liouville}

In this appendix, we provide the details of solving the Liouville equation~\eqref{eq:df_dt1}. 
Moving the first term on the right-hand side to the left, we can write
%----------------------------------------
\bea
&& e^{-i{\cal H}t}\left\{i\frac{\partial}{\partial t}\left(e^{i{\cal H}t} f e^{-i{\cal H}t}\right)\right\}e^{i{\cal H}t}=i\frac{\partial f}{\partial t}-[{\cal H},f]\nonumber\\
&& =[{\cal H},\rho_{l}]
-i\frac{\partial\rho_l}{\partial t}=-e^{-i{\cal H}t}\left\{i\frac{\partial}{\partial t}\left(e^{i{\cal H}t} \rho_l e^{-i{\cal H}t}\right)\right\}e^{i{\cal H}t}, \nonumber\\
\eea
%----------------------------------------
therefore
%----------------------------------------
\bea\label{eq:df_dt2}
i\frac{\partial}{\partial t}\left(e^{i{\cal H}t} f e^{-i{\cal H}t}\right)
=-i\frac{\partial}{\partial t}\left(e^{i{\cal H}t} e^{a(t)} e^{-i{\cal H}t}\right)
=-i\frac{d}{d t} e^{a_H(t)}, 
\hspace{-0.5cm}\nonumber\\
\eea
%----------------------------------------
where $a_H(t)=e^{i{\cal H}t}a(t)e^{-i{\cal H}t}$ is the Heisenberg representation of the operator $a(t)$ defined by Eq.~\eqref{eq:rho_l}.
Here, by full derivative $d/dt$ we mean the derivative with respect to both Heisenberg operators and external parameters on which these operators depend.

Next we will use the Duhamel formula
%----------------------------------------
\bea\label{eq:Duhamel}
\frac{d}{d t} e^{a_H(t)} = \int_0^1 ds\, e^{s a_H(t)}  \dot{a}_H(t) e^{(1-s) a_H(t)},
\eea
%----------------------------------------
where the operator  $\dot{a}_H(t)=d a_H(t)/d t$ does not commute with $a_H(t)$. 
As we will show below, $\dot{a}_H(t)$ is proportional to thermoelectric fields.
Therefore, within the linear response theory, 
the operator ${a}_H(t)$ in the exponential factors can be replaced by its equilibrium counterpart ${a}_H(t)\to \Omega_0-\beta K_0$. 
To proceed further we recall now that the time evolution of any operator $O_H(t)$ is given by $O_H(t)=e^{i{\cal H}t}O_H(0) e^{-i{\cal H}t}$,
therefore we have $O_H(t+\delta t)=e^{i{\cal H}(t+\delta t)}O_H(0)
e^{-i{\cal H}(t+\delta t)}=e^{i{\cal H}\delta t }O_H(t)e^{-i{\cal H}\delta t}$. Performing an analytic continuation 
$\delta t\to i\tau$ we obtain
%-----------------------------------------------
\bea\label{eq:op_heisenberg1}
O_H(t+i\tau)=e^{-\tau {\cal H}}
O_H(t)e^{\tau{\cal H}}.
\eea
%-----------------------------------------------
%from where we also obtain the relation
%-----------------------------------------------
%\bea\label{eq:op_heisenberg2}
%O_H(t+i\beta)\rho_0=e^{-\beta {\cal H}}O_H(t)e^{\beta {\cal H}}e^{\Omega_0-\beta K_0}=\rho_0 O_H(t).
%\eea 
%----------------------------------------------- 
Then, for the integrand of Eq.~\eqref{eq:Duhamel} we can write
%----------------------------------------
\bea\label{eq:Duhamel1}
&&e^{s a_H(t)} \dot{a}_H(t) e^{(1-s) a_H(t)}\nonumber\\
&&\simeq e^{-s \beta ({\cal H}_0-\mu{\cal N})} \dot{a}_H(t) e^{s\beta ({\cal H}_0-\mu{\cal N})} \rho_0
\nonumber\\
&&\simeq e^{s \alpha {\cal N}}\dot{a}_H(t+is\beta) e^{-s \alpha {\cal N}}\rho_0 =\dot{a}_H(t+is\beta) \rho_0.\qquad
\eea
%----------------------------------------
In the second line we approximated ${\cal H}_0\simeq {\cal H}$, and in the last step, we used the  fact that, as $\dot{a}_H(t)$ can be expressed by the operators of electric and heat currents (see below), it commutes with the total particle number operator ${\cal N}$.
Assuming now that the perturbation $\dot{a}_H(t)$ is switched on adiabatically from $t\to -\infty$, we can write 
%----------------------------------------
\bea\label{eq:Duhamel2}
\dot{a}_H(t+is\beta)&=&\int_{-\infty}^{t}
dt'\frac{d}{dt'}\dot{a}_H(t'+is\beta)\nonumber\\
&=& -\frac{i}{\beta} \int_{-\infty}^{t} dt' \frac{d}{ds}\dot{a}_H(t'+i\beta s),\quad
\eea
%----------------------------------------
therefore for Eq.~\eqref{eq:Duhamel} we obtain
%----------------------------------------
\bea\label{eq:Duhamel3}
-i\frac{d}{d t} e^{a_H(t)} &=&  -\frac{1}{\beta} \int_{-\infty}^{t} dt'\int_0^1 ds\,
\frac{d}{ds}
\dot{a}_H(t'+i\beta s)
\rho_0\nonumber\\
&=& -\frac{1}{\beta} \int_{-\infty}^{t} dt'\Big\{\dot{a}_H(t'+i\beta)-\dot{a}_H(t')\Big\}\rho_0\nonumber\\
&\simeq & 
 -\frac{1}{\beta} \int_{-\infty}^{t} dt'
\Big\{e^{-\beta {K}_0}\dot{a}_H(t')e^{\beta {K}_0}-\dot{a}_H(t')\Big\}\rho_0\nonumber\\
&=& -\frac{1}{\beta} \int_{-\infty}^{t} dt'
\Big\{\rho_0\dot{a}_H(t')-\dot{a}_H(t') \rho_0\Big\}\nonumber\\
&=& \frac{1}{\beta} \int_{-\infty}^{t} dt'
\left[\dot{a}_H(t'),\rho_0\right],\hspace{-1cm}
\eea
%----------------------------------------
where we again employed the relation~\eqref{eq:op_heisenberg1} and the commutation of $\dot{a}_H$ and ${\cal N}$.
Substituting now Eq.~\eqref{eq:Duhamel3} back into Eq.~\eqref{eq:df_dt2} we arrive at
%----------------------------------------
\bea
i\frac{\partial}{\partial t}\left(e^{i{\cal H}t}fe^{-i{\cal H}t}\right)
=\left[\frac{1}{\beta} \int_{-\infty}^{t} dt''\,\dot{a}_H(t''),\rho_0\right],
\eea
%----------------------------------------
and after integrating with the initial condition $f(-\infty)=0$ we obtain 
%----------------------------------------
\bea\label{eq:f_solution}
f(t)=-ie^{-i{\cal H}t}\left\{\int_{-\infty}^t\,dt' \left[\frac{1}{\beta} \int_{-\infty}^{t'} dt''\,\dot{a}_H(t''),\rho_0\right]\right\}e^{i{\cal H}t}.\nonumber\\
\eea
%----------------------------------------

Next we compute the time derivative of $a_H(t)$. From Eq.~\eqref{eq:rho_l} we can write explicitly 
%----------------------------------------
\bea\label{eq:a_H}
a_H(t)&=&\Omega_{l}(t)-A_H(t),\\
A_H(t)&=&\int d\bm r\Big[\beta(\bm r, t) T^{00}(\bm r,t)-\alpha(\bm r,
t)N^0(\bm r,t)\Big],
\nonumber\\
\eea
%----------------------------------------
where $\Omega_l(t)=-\ln\Tr {e^{-A_H(t)}}$.
In Eq.~\eqref{eq:a_H} we omit the index $H$ in the operators $T^{00}$ and $N^0$ for simplicity (the explicit time dependence already assumes the Heisenberg representation).
Note that, as the expression~\eqref{eq:f_solution} contains $\dot{a}_H(t)$ only in the commutator with $\rho_0$, it is sufficient to keep only the term $-\dot{A}_H(t)$ in $\dot{a}_H(t)$, as $\dot{\Omega}_l(t)$ is a number, and, therefore, commutes with any operator.
We next compute the time derivative of the operator $A_H(t)$, 
%----------------------------------------
\bea
\dot{A}_H(t)&\equiv& \frac{dA_H(t)}{dt}\nonumber\\
&=& \int d\bm r\Big[\partial_0\beta(\bm r, t)T^{00}(\bm r, t)+\beta(\bm r, t) \partial_0 T^{00}(\bm r, t)\nonumber\\
&-&\partial_0 \alpha(\bm r, t) N^0(\bm r,t)-\alpha (\bm r, t)\partial_0 N^0(\bm r, t)\Big].
\eea
%----------------------------------------
From the energy-momentum and particle number conservation laws ($J_\mu$ is the electric 4-current, and $F^{\mu\nu}$ is the 4-tensor of external electromagnetic field) we have
%----------------------------------------
\bea
\partial_\mu T^{\mu\nu}&=&\partial_0T^{0\nu}+\partial_iT^{i\nu}=J_\mu F^{\nu\mu},\\
\partial_\mu N^\mu&=&\partial_0N^0+\partial_iN^i=0,
\eea
%----------------------------------------
therefore, taking $\nu=0$, we obtain
$\partial_0T^{00}=-\partial_iT^{i0}-J_iE^{i}$, $\partial_0N^0=-\partial_i N^i$,
where $F^{0i}=-E^i$ is the electric field for fluids at rest.
Therefore [we omit the arguments $(\bm r,t)$ for simplicity],
%----------------------------------------
\bea\label{eq:dA_dt}
\dot{A}_H(t)
&=& \int d\bm r \Big[(\partial_i\beta) T^{i0}-(\partial_i\alpha) N^i
+(\partial_0\beta) T^{00}\nonumber\\
&-&(\partial_0\alpha) N^0 - \beta J_iE^{i}\Big],
\eea
%----------------------------------------
where we integrated by parts and dropped the surface term, which vanishes due to the boundary conditions at infinity. 
This expression can be further simplified by applying the Gibbs-Duhem relation for the grand canonical ensemble $dp=sdT+nd\mu$, 
where $p$ is pressure, $s$ is the entropy density, $T$ is the temperature, and $n$ is the particle number density. In terms of the variables $\beta=T^{-1}$ and $\alpha=\mu/T$ this relation reads 
%----------------------------------------
\bea\label{eq:Gibbs_Duhem}
\beta dp = -nhd\beta+nd\alpha,
\eea
%----------------------------------------
with $h=\mu+sT/n$ being the enthalpy per particle. As this relation holds for any quasiequilibrium variations of $p$, $\beta$, and $\alpha$, it can also be applied for the partial derivatives of these parameters.
Since we consider a fluid at rest, the gradient of pressure vanishes due to the condition of mechanical equilibrium $\partial_i p=0$.
Employing this, we arrive at $\partial_i\alpha=h\partial_i\beta$, and substituting this back into Eq.~\eqref{eq:dA_dt} we finally obtain
%----------------------------------------
\bea\label{eq:A_dot}
\dot{A}_H(t) &=&
\int d\bm r \Big[(\partial_0\beta) T^{00}-(\partial_0\alpha) N^0\nonumber\\
&+&(\partial_i\beta)H^i - \beta E_{i} J^i\Big],
\eea
%----------------------------------------
where we  defined the heat current operator by
%----------------------------------------
\bea\label{eq:heat_current}
 H^i(\bm r,t)=T^{i0}(\bm r,t)-hN^i(\bm r,t).
\eea
%----------------------------------------
Note that the heat current
\eqref{eq:heat_current} differs from the net energy flow by the
particle-convection term $\propto h$.
This definition of the heat current can be understood by means of the first law of thermodynamics:
${\delta q}=Tn {\delta s_1}={\delta \ep}-h{\delta n}$, where $s_1={s}/{n}$ is entropy per particle, ${\delta \ep}$ and ${\delta n}$ are any variations of energy density and particle number density, and $\delta q$ is the heat released irreversibly per unit volume due to these variations. (Note that the quantity ${\delta q'}=T{\delta s}={\delta \ep}-\mu {\delta n}$ includes also reversible entropy production $Ts_1\delta n$ due to particle convection as well, which should not be included in the definition of heat flux.)

Returning now to Eq.~\eqref{eq:f_solution} and substituting the expression~\eqref{eq:A_dot}, we obtain
%----------------------------------------
\bea\label{eq:ft_final}
f(t)=ie^{-i{\cal H}t}\Bigg\{\int_{-\infty}^t\,dt'
  \Big[V(t')+W(t')
+U(t'),\rho_0\Big]\Bigg\}e^{i{\cal H}t},\hspace{-0.5cm}\nonumber\\
\eea
%----------------------------------------
where we introduced the following notations
%----------------------------------------
\bea\label{eq:V_pert0}
V(t) &=& -\frac{1}{\beta} \int_{-\infty}^{t} dt'' \! \int d\bm
r\,\beta(\bm r,t'')J^i(\bm r, t'')E_{i}(\bm r, t''),
\nonumber\\\\
\label{eq:W_pert0}
W(t) &=& \frac{1}{\beta} \int_{-\infty}^{t} dt'' \! \int d\bm r\,H^i(\bm r,t'') \partial_i\beta(\bm r,t''),\\
\label{eq:U_pert0}
U(t) &=& \frac{1}{\beta} \int_{-\infty}^{t} dt'' \! \int d\bm
r\,\Big[\partial_0\beta(\bm r,t'')T^{00}(\bm r,t'')\nonumber\\
&-&\partial_0\alpha(\bm r,t'')N^0(\bm r,t'')\Big].
\eea
%----------------------------------------
Note that $U_H(t')$ contains the operators of the energy density $T^{00}$ and the net particle density $N^0$, which do not contribute to the averages of the current operators according to Curie’s theorem. The latter states that in an isotropic medium the correlations between operators of
different rank vanish~\cite{1963PhT....16e..70D,1963AmJPh..31..558D}. Therefore, in the solution~\eqref{eq:ft_final}, the term $U(t')$ can be dropped.  Within the linear response approximation, we can further ignore the spacetime dependence of $\beta(\bm r',t'')$ in the integrand of Eq.~\eqref{eq:V_pert0} and cancel it with the term $1/\beta$.

\section{Green's functions}
\label{app:Green_func}

Let us consider a generic transport coefficient given by the integral 
%----------------------------------------
\bea\label{eq:chi}
\chi(\omega) &=& 
i\!\int_{-\infty}^t dt' \!\int_{-\infty}^{t'} dt''
e^{i\omega^\star(t-t'')}\nonumber\\
&\times&\!\int d\bm r'' \langle\big[X(\bm r,t),Y(\bm r'', t'')\big]\rangle,
\eea
%----------------------------------------
where $X$ and $Y$ are any operators.
Taking into account that in an isotropic and homogeneous medium, the retarded Green's function in the integrand of \eqref{eq:chi} is a function only of the coordinate differences $\bm r-\bm r''$ and $t-t''$, and making variable changes $t'-t=t_1$, $t''-t=t_2$, $\bm r-\bm r'' =\bm r'$, we obtain
%----------------------------------------
\bea\label{eq:chi1}
\chi(\omega) &=& i\!\int_{-\infty}^t dt' \!\int_{-\infty}^{t'} dt'' e^{i\omega^\star(t-t'')} \nonumber\\
&\times& \!\int d\bm r'' \langle\big[X(\bm r-\bm r'',t-t''),Y(0)\big]\rangle\nonumber\\
&=& i\!\int_{-\infty}^0 dt_1 \!\int_{-\infty}^{t_1} dt_2
e^{-i\omega^\star t_2}\!
\nonumber\\
&\times&\int d\bm r' \langle\big[X(\bm r',-t_2),Y(0)\big]\rangle.
\eea
%----------------------------------------
Reversing the Fourier transformation of the retarded Green's function defined as
%----------------------------------------
\bea\label{eq:Pi_XY}
\Pi_{XY}^R(\omega)
= -i\!\int_{0}^{\infty} dt\, e^{i\omega t}\! \int d\bm r \langle \big[X(\bm r,t), Y(0)\big]\rangle, 
\eea
%----------------------------------------
we obtain
%-----------------------------------------------
\bea
-i\!\int d\bm r' \langle\big[X(\bm r',-t_2),Y(0)\big]\rangle=
\int_{-\infty}^{\infty} \frac{d\omega'}{2\pi}
e^{i\omega' t_2}\,\Pi^R_{XY}(\omega'),\hspace{-0.5cm}\nonumber\\
\eea
%-----------------------------------------------
therefore, substituting this into Eq.~\eqref{eq:chi1} and recalling that $\omega^\star=\omega+i0$, we obtain
%-----------------------------------------------
\bea\label{eq:chi2}
\chi (\omega)
&=& -\int_{-\infty}^0 dt_1 \! \int_{-\infty}^{t_1} dt_2\, e^{-i\omega^\star t_2}\!\int_{-\infty}^{\infty}\frac{d\omega'}{2\pi} e^{i\omega' t_2}\Pi^R_{\hat{X}\hat{Y}}(\omega')\nonumber\\
&=&-\int_{-\infty}^{\infty}\frac{d\omega'}{2\pi}\,\Pi^R_{\hat{X}\hat{Y}}(\omega')\int_{-\infty}^0 dt_1\! \int_{-\infty}^{t_1} dt_2\, e^{i(\omega'-\omega^\star) t_2}\nonumber\\
&=&-\int_{-\infty}^{\infty}\frac{d\omega'}{2\pi}\,\Pi^R_{\hat{X}\hat{Y}}(\omega')\int_{-\infty}^0 dt_1 \,
\frac{e^{i(\omega' -\omega-i0)t_1}}{i(\omega'-\omega-i0)}\nonumber\\
&=&i
\oint\frac{d\omega'}{2\pi i}\,
\frac{\Pi^R_{\hat{X}\hat{Y}}(\omega') }{(\omega' -\omega-i0)^2}\nonumber\\
&=&i\frac{d}{d\omega}
\oint\frac{d\omega'}{2\pi i}\,
\frac{\Pi^R_{\hat{X}\hat{Y}}(\omega') }{\omega' -\omega-i0}
= i\frac{d}{d\omega}\Pi^R_{XY}(\omega),
\eea
%-----------------------------------------------
where we closed the integral in the upper-half plane
where the retarded Green's function is analytic, and applied Cauchy's integral formula.

Now from Eq.~\eqref{eq:Pi_XY} we find that
%-----------------------------------------------
\bea\label{eq:parity_green}
\big\{\Pi^R_{XY}(\omega)\big\}^* =
\Pi^R_{XY}(-\omega),
\eea
%-----------------------------------------------
which is due to the fact that the commutator of two Hermitian (current) operators is anti-Hermitian
%-----------------------------------------------
\bea
\big[X(\bm r, t) Y(0)\big]^\dagger 
&=& Y(0)^\dagger X(\bm r, t)^\dagger-X(\bm r, t)^\dagger Y(0)^\dagger\nonumber\\
&=&
\big[Y(0)X(\bm r, t)\big]
=-\big[X(\bm r, t)Y(0)\big].\nonumber\\
\eea
%-----------------------------------------------
As a result, the eigenvalues, and, therefore, also the thermal average of this commutator, are imaginary, which was used in Eq.~\eqref{eq:parity_green}.
Splitting $\Pi^R_{XY}(\omega)$ in real and imaginary parts we obtain from Eq.~\eqref{eq:parity_green}
%-----------------------------------------------
\bea\label{eq:parity_green1}
{\rm Re}\Pi^R_{XY}(-\omega)&=&
{\rm Re}\Pi^R_{XY}(\omega),\\
{\rm Im}\Pi^R_{XY}(-\omega)&=&
-{\rm Im}\Pi^R_{XY}(\omega),
\eea
%-----------------------------------------------
\ie, the real and imaginary parts of the Green's function are, respectively, even and odd functions of the frequency.
Taking this into account, from Eq.~\eqref{eq:chi2} we find that in  the limit of vanishing frequency $\omega \rightarrow 0$ $\chi$ is real and is given by
%-----------------------------------------------
\bea\label{eq:chi4}
\chi
=-\frac{d}{d\omega}{\rm Im}\Pi^R_{XY}(\omega)\bigg\vert_{\omega=0},
\eea
%-----------------------------------------------
which was used to write down the Kubo formulas~\eqref{eq:sigma}--\eqref{eq:Qsigma}.
%\end{widetext}

\bibliographystyle{apsrev4-2}
\bibliography{KuboNJL.bib}{}

urr
\end{document}